\documentclass[a4paper,fleqn,useAMS,usenatbib]{mnras}

\pdfoutput=1

\usepackage{natbib}
\usepackage{mathptmx}

\usepackage[T1]{fontenc}
\usepackage{ae,aecompl}

\usepackage[dvips]{graphicx}
\usepackage{amsmath}
\usepackage{amsfonts}
\usepackage{amssymb}
\usepackage{comment}
\usepackage{bm} 
\usepackage[dvipsnames]{xcolor}
\usepackage[]{hyperref}
\hypersetup{
	colorlinks=true,	
	urlcolor=MidnightBlue,		
	hypertexnames=true,
	plainpages=false,
	naturalnames=true,
	pdftitle={The Multiplicity Distribution of Kepler's Exoplanets},    
	pdfauthor={Cool Worlds Lab},     
	linkcolor=WildStrawberry,          
	citecolor=ForestGreen,        
}

\usepackage{amsmath}
\usepackage{amsfonts}
\usepackage{amssymb}
\usepackage{wasysym}
\usepackage[percent]{overpic}

\makeatletter
\def\env@matrix{\hskip -\arraycolsep 
  \let\@ifnextchar\new@ifnextchar
  \array{*{\c@MaxMatrixCols}c}}
\makeatother

\title[Out-of-Transit Effects in Eccentric Systems]{Higher Order Harmonics in the Light Curves of Eccentric Planetary Systems}
\author[Z. Penoyre \& E. Sandford]{Zephyr Penoyre$^{1}$\thanks{E-mail:
\href{mailto:zpenoyre@ast.cam.ac.uk}{zpenoyre@ast.cam.ac.uk}} and Emily Sandford$^{2}$ \\
$^{1}$Institute of Astronomy, University of Cambridge, Madingley Road, Cambridge, CB3 0HA, United Kingdom \\
$^{2}$Dept. of Astronomy, Columbia University, 550 W. 120th Street, New York NY 10027, USA \\
}

\date{Published in MNRAS}

\pubyear{2019}

\begin{document}
\label{firstpage}
\pagerange{\pageref{firstpage}--\pageref{lastpage}}
\maketitle

\begin{abstract}
As a planet orbits, it causes periodic modulations in the light curve of its host star. Due to the combined effects of the planet raising tides on the host star, relativistic beaming of the starlight, and reflection of light off the planet's surface, these modulations occur at the planet's orbital frequency, as well as integer multiples of this frequency. In particular, planets on eccentric orbits induce third and higher-order harmonics in the stellar light curve which cannot be explained by circular-orbit models. Even at moderate eccentricities, such as those typical of Solar System planets, these harmonics are detectable in current and future photometric data. We present an analysis of the harmonics caused by tides, beaming, and reflection in eccentric planetary systems. We explore the dependence of these signals on the parameters of the system, and we discuss prospects for current and future observations of these signals, particularly by the NASA TESS mission. Finally, we present publicly available code for computation of light curves with tidal, beaming, and reflection signals, \texttt{OoT}.
\end{abstract}


\section{Introduction}


A planet may influence its star's light curve in two ways: either directly, by exerting gravitational influence on the star and changing its pattern of emission, or indirectly, by leaving the star itself unaffected but redirecting its emitted light. Examples of direct influence include light curve modulations due to tides raised on the star by an orbiting planet, as well as relativistic beaming of the starlight as the star orbits the planet-star barycenter. Indirect influences include reflection of starlight off the planet's surface or atmosphere, as well as planetary transits. 

Of these four, transits (see e.g. \citealt{Charbonneau00}) are best understood and most studied because of their unambiguous signal in the observed light curve and its straightforward geometric interpretation. However, we only observe transits of planets whose orbits are inclined nearly edge-on to our line-of-sight, which comprise a small fraction of all planetary systems.

In contrast, the three \textit{out-of-transit} (OoT) signals listed above---tides, beaming, and reflection---are completely general. Indeed, all three signals should be present in every planetary system (though they may be vanishingly small). Furthermore, like transits, they are well constrained: simple analytic models can be constructed to model each with few free parameters and assumptions about the planet or the star (e.g. the circular-orbit models of \citealt{Faigler11} and \citealt{Shporer17}). 

In this paper, we investigate the effects of these three signals on the light curves and power spectra of planet-hosting stars, for the general case of planets on eccentric orbits. Much previous work has been done on these signals independently, on the extreme (and less analytically tractable) case of binary star systems, and on systems confined to perfectly circular orbits. 

Here, we build upon the simple analytic framework for eccentric planetary tides presented in \citep{Penoyre18} (PS18), as well as correspondingly simple analytic models for beaming and reflection from the literature \citep{Loeb03, Charbonneau99}, to explore the signals that these effects will generate in the power spectra of stellar light curves. In section~\ref{extendedIntro}, we contextualise this work among previous investigations into tides, both observed and theoretical. In section~\ref{photometric}, we present the simple analytic models we use to calculate the effects of tides, beaming, and reflection in stellar light curves. In section~\ref{powerSpectra}, we show the light curves and power spectra of these three signals and investigate their observable signatures. In section~\ref{prospects}, we consider the detectability of these effects, particularly in power spectra of light curves from the upcoming NASA TESS mission, and explore how these signals change across parameter space. We conclude in section~\ref{discussion}, and present code to calculate these signals, the \texttt{OoT} (short for ``out-of-transit") package, in Appendix~\ref{code}.

\section{Previous work on tides raised by planets}\label{extendedIntro}

Tides raised on stars by orbiting companions, and their resulting photometric signatures, have been widely studied for many decades. The early definitive theoretical text on this subject is \cite{Kopal59}, which lays out a mathematical framework for ``ellipsoidal variable" binary star systems. \cite{Morris85} presents simplified expressions and an observational catalog of such systems. \cite{Kumar95} consider a more general excitation of oscillations in eccentric binaries. Recent work on tides in binary star systems includes the investigation of equilibrium and dynamical tides in eccentric ``heartbeat" binaries \citep{Fuller17}.

In the context of the results from recent photometric surveys, including NASA's \textit{Kepler} mission, it is necessary to consider this existing literature in a new light and a new parameter space. A binary companion raising tides on a star need not be another star---it could be a planet. Indeed, \cite{Pfahl08}, before the launch of \textit{Kepler}, predicted the expected light curve signal of tides raised by planets.

Tidal photometric signals due to planets are smaller than those due to stars, but PS18 demonstrate that, particularly for massive planets on eccentric orbits which pass close to their host stars, many more such signals could be detectable than are currently known. \textit{Kepler} has revealed many good candidates for this analysis, including a population of eccentric, hot, Jupiter-mass planets (see \citealt{Kane12b}, or \citealt{Winn15} for a review). Radial velocity surveys, too, find many eccentric planets---for example, \cite{Wittenmyer17}, in figure 4, plot the eccentricity of 116 confirmed planets orbiting giant stars as a function of their periastron distance; the majority have eccentricities greater than 0.1, even for extremely close-in orbits. PS18 calculate the functional form of the tidal signal in the light curve; in this work, we extend their investigation to the tidal signal in the power spectrum, and consider it alongside the other effects of beaming and reflection.

Another thread in the existing literature approaches the question of planetary tides from the observational side. Ellipsoidal variations have been commonly observed in the light curves of transiting planets, beginning with HAT-P-7 b \citep{Welsh10}. A number of papers have addressed the question of detecting and fitting these light curve signals in order to better constrain the star and planet properties, both for planetary systems in general (see below) and for interesting individual cases (e.g. \citealt{LilloBox14}, who adopt the functional forms derived by \citealt{Pfahl08} to fit the light curve of Kepler-91 b, an eccentric hot Jupiter orbiting a giant star).

\cite{Faigler11} introduce the BEER algorithm (BEaming, Ellipsoidal variations, and Reflections) to find non-transiting planetary companions to observed stars, but do not consider planets on eccentric orbits. \cite{Jackson12} offer a semi-analytic model for planetary tides for use in fitting light curves, but again do not consider eccentric planets.

Several attempts have been made to include the effects of eccentricity in phase curve models, including the work of \cite{Kane12} and \cite{Placek14}. \cite{Gai18} collate and compare these models (as well as some circular-orbit models, e.g. \citealt{Jackson12}) for the same input planet parameters and find significant variation in their predicted light curves, indicating that there is some confusion in the literature as to the correct model. PS18 address this confusion by presenting a purely analytic formulation for tides due to eccentric planets, and in this work, we adopt their model for our investigation into the power spectra of tidal signals.

Finally, planetary tides have been detected in the power spectra of certain \textit{Kepler} planets, although not recognised as such. In particular, oscillations at 3 times the transiting companion's orbital frequency have been noted in the power spectra of a number of \textit{Kepler} light curves \citep{Esteves13,Armstrong15,Cowan17}. We demonstrate here that this pervasive quirk can in fact be a natural consequence of eccentric planetary orbits.

\section{Photometric effects of tides, beaming, and reflections}
\label{photometric}

Throughout this work, we are primarily concerned with the fractional change in luminosity of a star, as a function of time, due to the presence of a planetary companion. We will start by summarising the effects of tides, beaming, and reflections on a star's light curve.

We shall not discuss transits (eclipses of a star by an orbiting planet), though they too have non-trivial power spectra, as they are already well understood (see e.g. \citealt{Seager03}), and only a small fraction of planets will have sufficiently fortuitous alignments for an eclipse to be visible. Even in systems with visible transits, the below analysis can be performed on the rest of the light curve. Then new properties of the system can be derived, or tighter constraints placed on those derived from the transit.

Let us define the fractional change in luminosity
\begin{equation}
\delta = \frac{\Delta L}{L},
\end{equation}
where $L$ is the luminosity the star would have in the absence of a companion, and $\Delta L$ is the apparent change in this luminosity. We will frequently make use of the subscripts $t,\ b,\ r,$ and $\Sigma$ to specify the tidal, beaming, reflection, and total effects respectively.

Here, we consider a planet with mass $M_p$ and radius $R_p$ on an orbit with semi-major axis $a$ and eccentricity $e$ around a star with mass $M$ (assumed to be $\gg M_p$) and radius $R$.

The orbit obeys
\begin{equation}
\label{eq:r}
r(t) = \frac{a(1-e^2)}{(1+e \cos\Phi)} = a(1-e \cos \eta),
\end{equation}
where $r$ is the orbital radius and $\Phi(t)$ is the angle, relative to the star, between the planet's position at periapse and its position at time $t$ (often called the true anomaly). The eccentric anomaly, $\eta(t)$, is a useful simplification satisfying
\begin{equation}
\label{eta}
\sqrt{1+e} \tan \frac{\eta}{2} = \sqrt{1-e} \tan \frac{\Phi}{2}.
\end{equation}
$\eta(t)$ can also be found (numerically) from
\begin{equation}
\label{tEta}
t(\eta) = \sqrt{\frac{a^3}{GM}} (\eta - e \sin \eta),
\end{equation}
and thus the position of the planet can be found at any given $t$
\citep{Binney08}.

From this we can read off the period
\begin{equation}
P=2\pi \sqrt{\frac{a^3}{GM}}
\end{equation}
and the planet's orbital frequency
\begin{equation}
\omega_p = \frac{2\pi}{P} = \sqrt{\frac{GM}{a^3}}.
\end{equation}

All of the signals discussed in this paper will be composed of oscillations with a frequency equal to integer multiples of $\omega_p$. We shall often make use of the shorthand of saying such a signal of frequency $\omega$ is at the $n^{th}$ \textit{harmonic}, where $\omega = n \omega_p$. This makes out-of-transits easy to distinguish from periodic signals in the power spectrum related to stellar activity (e.g. starspots), which occur at the stellar rotation frequency (e.g. \citealt{McQuillan14}).

We will work in spherical co-ordinates with polar angle $\theta$ (ranging from 0 at the north pole, to $\pi$ at the south pole) and azimuthal angle $\phi$ ranging from 0 to $2\pi$. We orientate the system such that the planet orbits in the equatorial plane ($\theta=\frac{\pi}{2}$) and $\phi=0$ points towards the position of the planet's periapse. It will be convenient to define a second azimuthal angle, relative to the planet's position at time $t$,
\begin{equation}
\psi(t,\phi) = \phi - \Phi(t).
\end{equation}

The observed light curve and power spectrum depend on the orientation of the system relative to the observer. Let the observer be situated at some angle ($\theta_v,\phi_v$), where $\theta_v=0$ is equivalent to viewing the system from the top, or face-on, and $\theta_v=\frac{\pi}{2}$ corresponds to an edge-on view (in which transits would be visible). We define a second azimuthal viewing angle, relative to the planet's position:
\begin{equation}
\psi_v(t) = \phi_v - \Phi(t).
\end{equation}

Four example choices of viewing angle ($\theta_v,\phi_v$) are sketched in the diagrams accompanying Table~\ref{viewingAngles}. With these co-ordinates in hand, we can calculate $\delta$ for tides, beaming, and reflections.

\subsection{Tides}\label{subsec:tides}

Following the derivation in PS18, for a non-rotating star tidally distorted by a small perturber, the deviations to the stellar radius at the surface can be described by
\begin{equation}
\frac{\Delta R(t,\theta,\phi)}{R} = \beta\frac{M_p}{M}\left(\frac{R}{r(t)}\right)^3 \frac{3 \sin^2 \theta \cos^2 (\psi(t,\phi)) - 1}{2}
\label{eq:tides}
\end{equation}
where $\beta$ is a dimensionless constant describing the response of the star to tides. For simplicity we take the result for a stellar surface that follows the equipotential, $\beta=1$\footnote{$\beta \simeq 1$ for most stars. $\beta$ increases for shallower internal density profiles, with a maximum of 2.5 for a uniform-density star (Generozov et al., in prep).}.

This form assumes that the tide is well represented by its quadrupole moment, but higher order tidal perturbations may have some small contribution. The amplitude of these distortions is reduced by a factor of roughly $R/r$. We ignore this effect here, though it can be shown \citep{Morris93} that these higher order modes will cause light curve modulations at various harmonics.

The tidal distortions described by Equation~\ref{eq:tides} affect the light curve in two ways: (i) by changing the apparent sky-projected area of the star as the planet orbits and (ii) by changing the gravitational force (and hence the pressure, temperature and flux) at the surface. These effects add together; the star is elongated towards and away from the planet, meaning that the sides facing towards and away from the planet are both smaller (by effect (i)) and dimmer (by effect (ii)). 

As the planet completes one orbit, we see, in turn, the dimmer side of the star facing the planet; the first bright side; the dimmer side of the star facing away from the planet; and the other bright side. We therefore observe a light curve signal at the second harmonic of the planetary frequency $\omega = 2\omega_p$. This is true even for circular-orbit planets.

The total photometric change, integrated over the stellar surface visible from ($\theta_v,\phi_v$), can be shown analytically to be
\begin{equation}
\label{deltaTides}
\delta_t(t,\theta_v,\phi_v) = - \frac{13}{16}\frac{M_p}{M}\left(\frac{R}{r}\right)^3 (3 \sin^2 \theta_v \cos^2 \psi_v - 1)
\end{equation}
(PS18, section 3.2). 
Here we have used the Eddington limb darkening law,
\begin{equation}
\frac{I(\mathbf{r})}{I_0} = \frac{2}{5} \left( 1+\frac{3}{2}(\mathbf{\hat{r}}\cdot \mathbf{\hat{l}}) \right)
\end{equation}
where $I$ is the intensity at some point $\mathbf{r}$ on the surface of the star, $\mathbf{l}$ is the line of sight direction (hats denote unit vectors), and $I_0$ is the intensity at the projected centre. This law gives a good general fit to most systems and is sufficiently accurate for calculations integrating over the whole area of the star. It can fail at the very edge of the star but unlike for transit observations (particularly ingress and egress) the contribution from this region is negligible.

\subsection{Beaming}

\citet{Loeb03} give a simple expression relating the velocity of a star to its fractional change in luminosity due to relativistic beaming, using the spectral dependence of a typical Kepler star:
\begin{equation}
\delta_b \approx \frac{4 v_l}{c}.
\end{equation}
$c$ here is the speed of light, and $v_l$ is the projection of the star's orbital velocity along the line of sight. Assuming $M_p \ll M$, $v_l$ is equal to
\begin{equation}
v_{l}(t,\theta_v,\phi_v) = v_c\frac{M_p}{M} \sin\theta_v \left(\sin(\psi_v(t)) + e \sin\phi_v\right),
\end{equation}
where $v_c$ is the characteristic velocity of the planet,
\begin{equation}
v_c = \sqrt{\frac{GM}{a(1-e^2)}}
\end{equation}
\citep{Lovis10}.

Thus we find
\begin{equation}
\delta_b(t,\theta_v,\phi_v) \approx \frac{4}{c} \sqrt{\frac{GM}{a(1-e^2)}} \frac{M_p}{M} \sin\theta_v \left(\sin\psi_v + e \sin\phi_v\right).
\label{deltaBeaming}
\end{equation}

\subsection{Reflection}

Of the processes considered in this work, the reflection of a star's light by an orbiting planet requires the largest degree of approximation. The fraction of the light incident on the planet which is reflected (the albedo) depends greatly on the properties of the planet's atmosphere or surface. The chemical composition, thermodynamic properties, and even the weather systems in a planetary atmosphere can affect the amount of light reflected (see e.g. \citealt{Jansen17,Cowan11}).

Here we will use one of the simplest relevant models: a perfectly scattering surface (often called a Lambert surface/sphere) which absorbs radiation and re-emits some fraction of it isotropically. 

In this model, an infinitesimal surface element absorbs energy at a rate proportional to the flux from the star and the apparent surface area of the element (taking into account inclination). It then radiates some fraction, $A_b$ (the Bond albedo), of this energy out uniformly over a solid angle of $2\pi$.

This gives an observed apparent luminosity that satisfies
\begin{equation}
\delta_r(t,\theta_v,\phi_v) = \frac{A_g}{\pi} \left(\frac{R_p}{r}\right)^2 (\sin \gamma + (\pi - \gamma) \cos \gamma)
\label{deltaReflection}
\end{equation}
where $\gamma$ is the angle between the line of sight and the direction from which the planet is illuminated. Thus $0<\gamma<\pi$ and $\cos{\gamma} = -\sin{\theta_v} \cos{\psi_v}$. $A_g$ is the geometric albedo, and is equal to $\frac{2}{3}A_b$ for a Lambert sphere. (See \citealt{Seager10} for a more detailed derivation of these results.)

We do not consider thermal radiation from the planet, i.e. energy from the star re-radiated by the planet at its equilibrium temperature. As even the closest-in planets have equilibrium temperatures of $\sim 1000 \mathrm{K}$ or below, the contribution to the light curve when viewed in wavelengths close to optical light (as is true for the \textit{Kepler} and TESS surveys) will be small.

\begin{figure}
\includegraphics[width=\columnwidth]{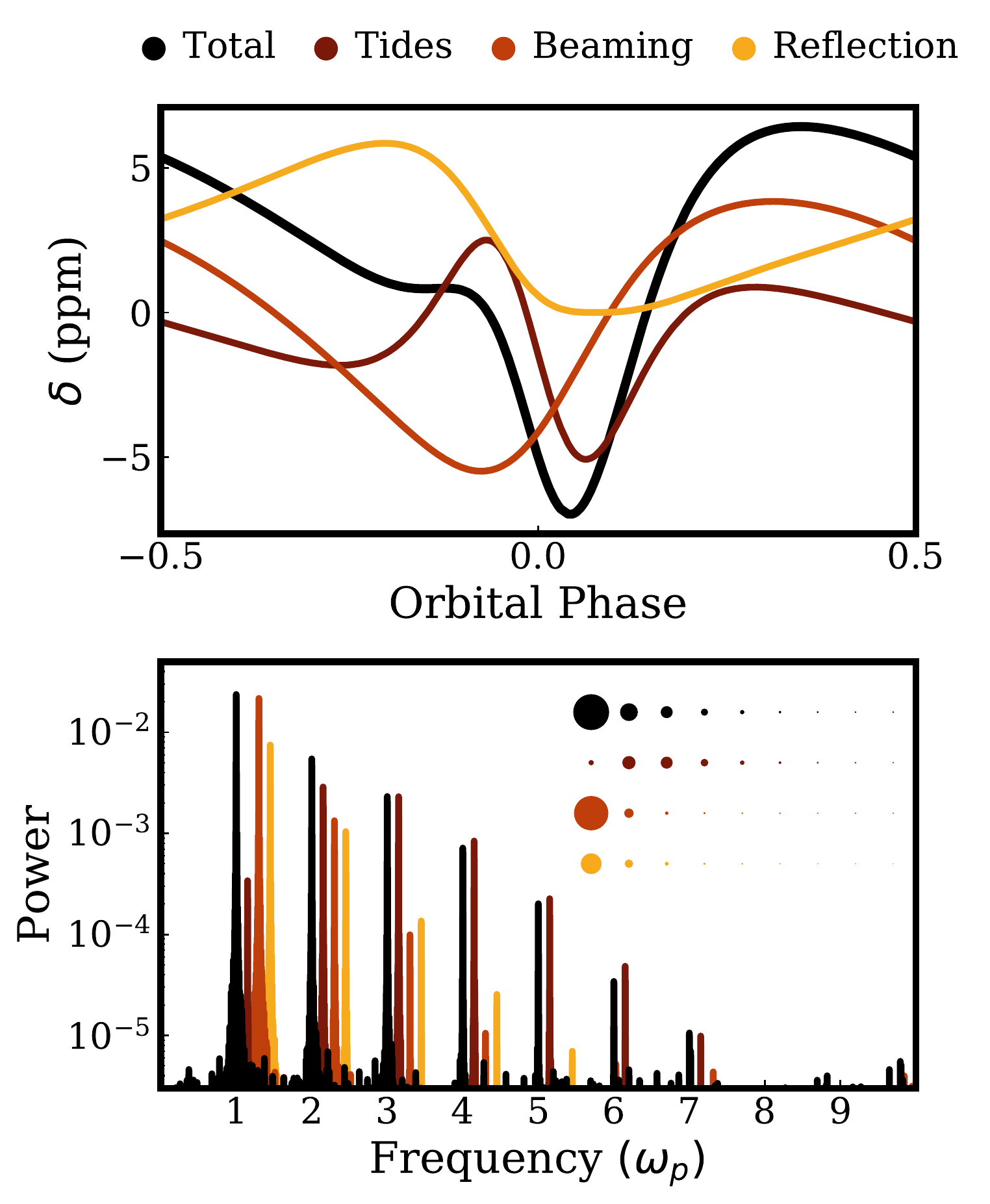}
\caption{The light curve (top) and power spectrum (bottom) caused by a planet with $M_p=3 M_J$ and $R_p = 1.5 R_J$ on an orbit with $a=0.07 AU$ and $e=0.25$ around a star with $M=1 M_\odot, R=1.2 R_\odot$. The period of this planet is roughly 10 days, and we have set the orbital phase equal to 0 at the moment of periapse. The photometric contribution is shown individually for tidal distortions of the star (dark red), relativistic beaming caused by the star's motion (brick red) and reflections of the star's light by the planet (yellow), and the sum of these effects is shown in black. The power spectrum is computed for each individual signal over 100,000 data points and is normalised relative to the total signal. The power spectra have been shifted to make each visible (but all fall at integer multiples of $\omega_p$ in reality). The circles shown in top right give a simple visual representation of the power in each harmonic, with area proportional to power. The system is viewed from an angle in the equatorial plane of the orbit $(\theta_v = \frac{\pi}{2},\ \phi_v = \frac{\pi}{4})$.}
\label{intro}
\end{figure}

\section{Light Curves and Power Spectra}\label{powerSpectra}

Let us consider an example planet, with mass $M_p=3 M_J$ (Jupiter masses), radius $R_p = 1.5 R_J$ (Jupiter radii), and geometric albedo $A_g = 0.15$, on an orbit with semi-major axis $a=0.07 \mathrm{AU}$ and eccentricity $e=0.25$. Let it be the companion of a star with stellar mass $M=1\,M_\odot$ and radius $R=1.2\,R_\odot$. These properties have been chosen so as to make the contribution of tides, beaming and reflections approximately equal in magnitude, and we will use this example planet throughout. 

In PS18, we showed that such a planet is well within the distribution of confirmed exoplanets, and many systems exist which should show significantly larger photometric variation. As of writing,\footnote{Accessed 27 February 2019.} there are 30 confirmed exoplanets listed on the NASA Exoplanet Archive \citep{Akeson13} with $a<0.1\,\mathrm{AU}$, $e>0.1$ and $M_p > 1\,M_J$ (14 of which only have upper limits on eccentricity). Examples of such extreme planets can be found in \citet{Bonomo15,Stassun17} and 
\citet{Wittenmyer17}. In particular, both HAT-P-2 b \citep{deWit17} and HATS-41 b \citep{Bento18} are more extreme than our example planet in every respect (smaller $a$, larger $e$, larger $M_p$).

For any planet, the values of $\delta$ are small ($\ll 1$), due to the mass and radius of a planet being a fraction of that of a star. Thus the effects of tides, beaming, and reflection are independent and additive, so we can find each separately and add them to yield
\begin{equation}
\delta_\Sigma = \delta_t + \delta_b + \delta_r.
\end{equation}

Computing the power spectrum of these photometric variations is a more nuanced task, and so we fall back upon established tools and methods. For all power spectra shown here, we use the \texttt{LombScargle} function from the \texttt{astropy} package, which implements the methods of \citet{Press89} (see \citealt{Vanderplas17} for a full summary).

There is always a fear of artificially injecting a periodic signal into data through its processing. To avoid that in this work, we calculate $\delta$ at random points in time over an irrational multiple of the planet's orbital period. Unless otherwise specified, 10,000 data points are used, over an observational baseline of roughly 100 orbital periods.

Figure \ref{intro} shows the light curve, and the power spectrum, of our example planet (viewed edge-on, $\theta_v=\frac{\pi}{2}$, with $\phi_v=\frac{\pi}{4}$). The profiles have been calculated via \texttt{OoT}, a publicly available code that we are releasing alongside this work (see Appendix \ref{code}). The independent contributions of tides, beaming, and reflections are shown, as well as the total signal.

Looking first at the light curve, we see that the tidal, beaming, and reflection signals are all markedly different, and while each independently has a relatively simple shape, the combined signal does not. At some phases, the three add together to yield a large peak, whereas at others they cancel.

A perceptive reader may notice that only the beaming signal is centred around $\delta=0$. Reflections can only add to the observed luminosity of the star: some fraction of the light that would otherwise be lost is instead redirected to the observer. The tidal signal can be centred around positive or negative $\delta$ depending on viewing angle.

Moving to the power spectrum, we see not only the familiar peaks at $\omega_p$ and $2\omega_p$ (which are expected even for planets on circular orbits; see section~\ref{subsec:tides}), but also many peaks of similar amplitude reaching up to high harmonics. The peaks are discrete, and the power spectrum appears not as a continuous function but as a series of Dirac delta functions. Thus, the spectrum could be well represented by a Fourier series expansion.

The individual power spectra of the tidal, beaming, and reflection signals are also shown, and we can see that all three show some power in the $3^{\mathrm{rd}}$ harmonic and higher. The dominant contribution to the beaming and tidal signatures is still at the planet's orbital frequency, with the beaming signal falling off quickly at higher frequencies and the reflection signal falling off more slowly and extending to higher harmonics. 

The tidal signal is most interesting: a series of spikes that rises to a crest at some harmonic and then drops off as a long tail. For this example planet, the dominant contribution is still at the second harmonic (as expected for the circular-orbit case), though as we will show, this is not true for high eccentricities. There is also some contribution from tides at the planet's orbital frequency. 

\subsection{The role of viewing angle and eccentricity}

\begin{figure*}
    \begin{center}
    \includegraphics[width=\textwidth]{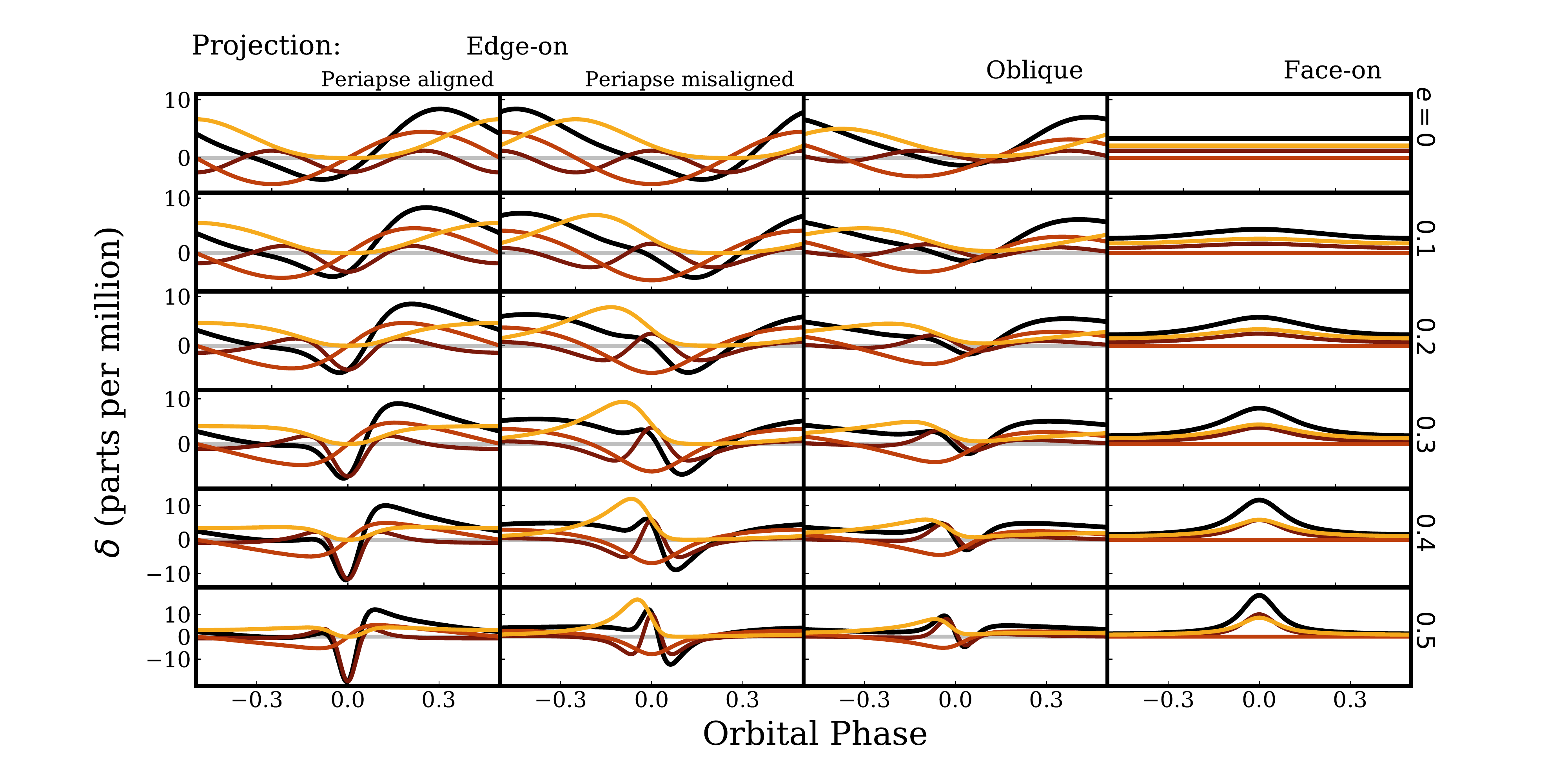} 
    
    \includegraphics[width=\textwidth]{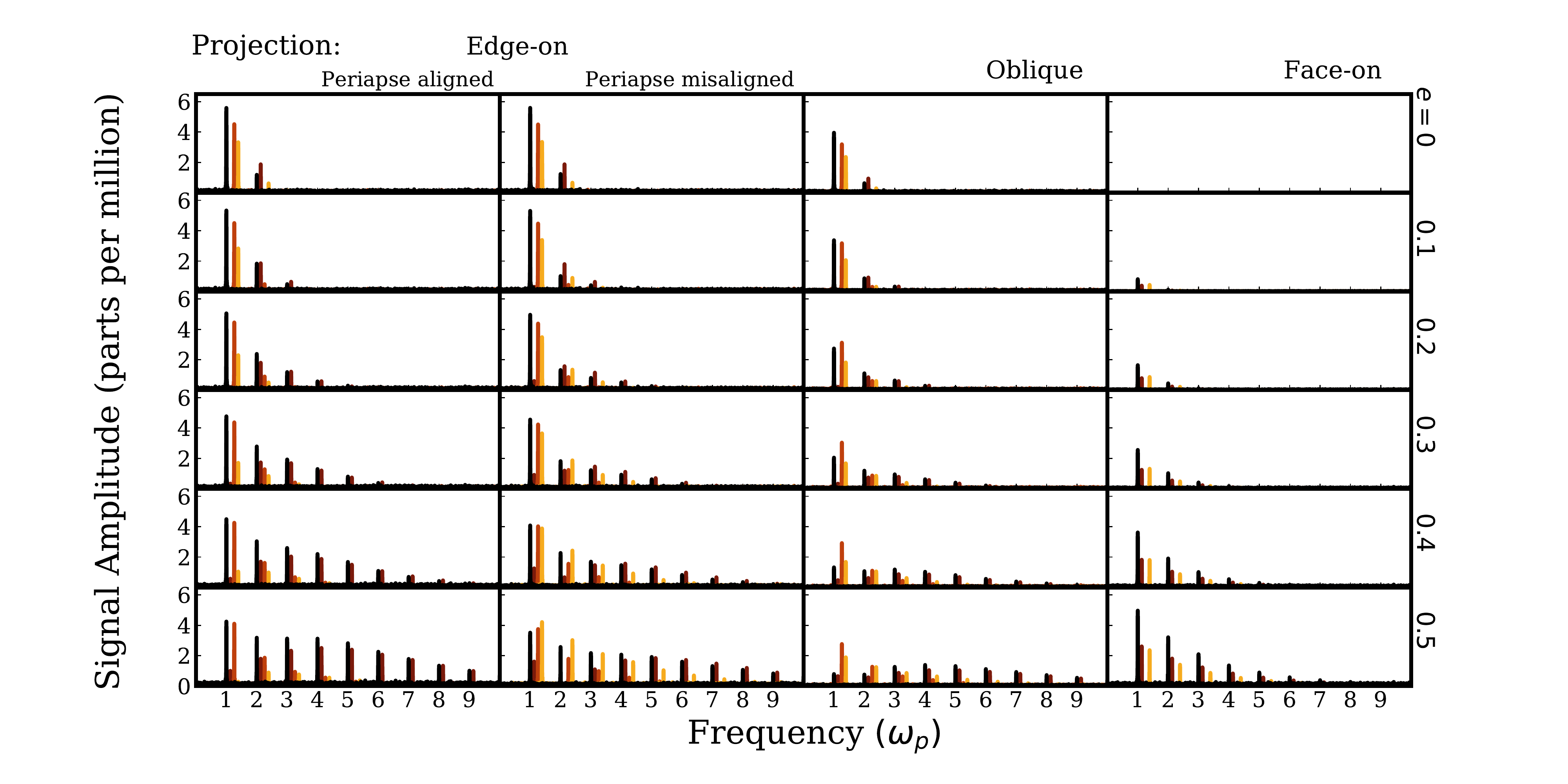}
    \end{center}
    \caption{Similar to Figure \ref{intro}, but shown for a variety of eccentricities and viewing angles (detailed in Table~\ref{viewingAngles}). We show independently the out-of-transit signal due to tides (dark red), beaming (brick red) and reflections (yellow), as well as the total signal (black). The same system properties, save for the varying eccentricity, are used in each row of subplots. Note that the y-scaling varies between rows in panel (a). The signal amplitude used in panel (b) is effectively equal to the coefficients of the Fourier series that can be used to construct $\delta(t)$.}
    \label{twoFig}
\end{figure*}

Moving to a wider array of possible system parameters, Figure~\ref{twoFig} shows similar light curves and power spectra for the same planet-star system, but now with varying eccentricity and viewing angle (defined in Table~\ref{viewingAngles}).

Examining first the light curves in Figure~\ref{twoFig}, we see that for $e=0$, each signal is well-represented by a sinusoid (perfectly so for tides and beaming). As we move to higher eccentricity, the signals develop richer features, the tidal signature comes to dominate, and the majority of the variation is now centred around the short window when the planet is near periapse. For approximately circular orbits, the signal drops off as the observer moves out of the orbital plane (through oblique to face-on projections); however, as shown in PS18, for eccentric systems, the variation in luminosity is almost equally visible from all angles (with both tides and, to a smaller extent, reflections being visible in perfectly face-on systems).

The power spectra in Figure~\ref{twoFig} encompass a lot of detail, so we shall go through them piece by piece. Here, we have expressed the spectra in terms of the signal amplitude
\begin{equation}
A(\omega) = 2\sqrt{\frac{P(\omega)}{N}}
\end{equation}
where $P(\omega)$ is the power and $N$ the number of data points. For sharply peaked spectra such as these, this is equivalent to the magnitude of the coefficients of the Fourier series representation of $\delta(t)$. Thus we can directly relate the amplitude of the power spectra to that of the light curves.

For planets on circular orbits, we see that the power spectrum is independent of $\phi_v$, and only the total amplitude depends on $\theta_v$ (this is not true at higher eccentricities). We also see that the reflection signal, even for circular orbits, is not perfectly represented by a single sinusoid, there is some small contribution at the second harmonic (which might otherwise be mistaken for a small tidal signal).

For small eccentricities, roughly equivalent to those of planets in our own solar system ($0.01 < e < 0.2$), we start to see higher-order harmonics.

As we move to larger eccentricities, the amplitude of higher-order harmonics, particularly those associated with tides, increases. A visible signal in the power spectrum appears for the face-on case. Only for these higher eccentricities does the dependence on the azimuthal viewing angle, $\phi_v$, become easily visible.

Note that the characteristic amplitude of individual peaks does not vary greatly. For highly eccentric systems, with significantly larger $\delta$, it is not the characteristic amplitude of the peaks but the number of harmonics that yields a larger signal in the light curve.

\begin{table}
\centering
\begin{tabular}{ c| c c c c } 
& \textbf{Edge-on} & & \textbf{Oblique} & \textbf{Face-on} \\
  & Periapse- & Periapse- &  & \\
  & Aligned & Misaligned &  & \\ 
 \hline
 \hline
 $\theta_v$ & $\frac{\pi}{2}$ & $\frac{\pi}{2}$ & $\frac{\pi}{3}$ & $0$ \\ 
 $\phi_v$ & $0$ & $\frac{\pi}{2}$ & $\frac{\pi}{4}$ & 0  \\
  & \includegraphics[width=0.18\columnwidth]{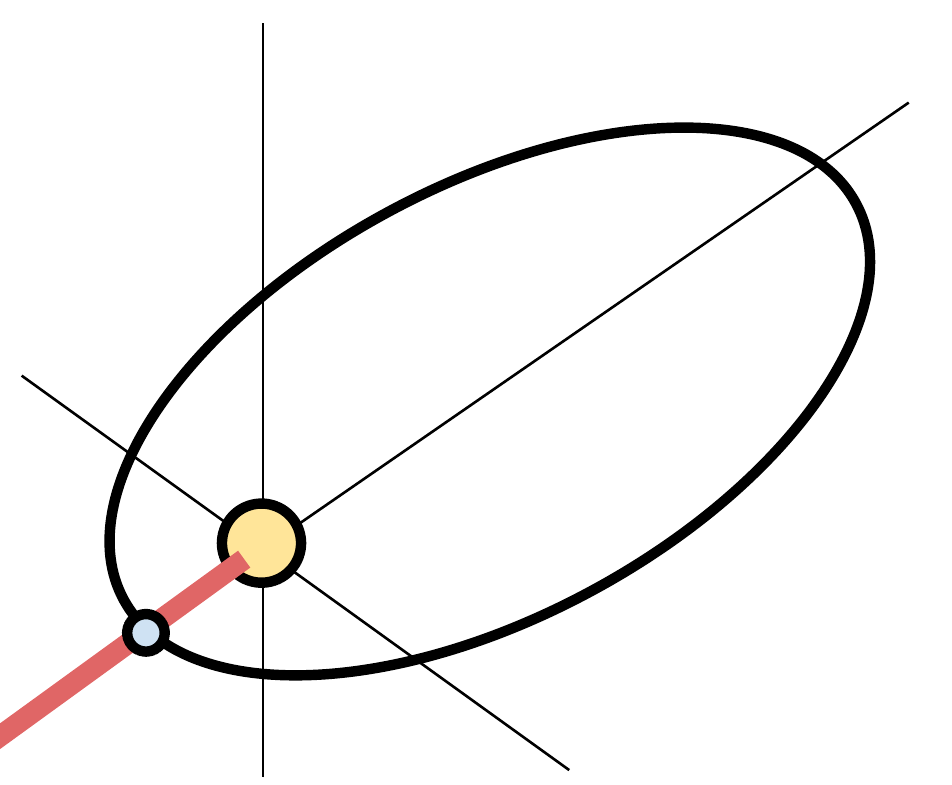} & \includegraphics[width=0.18\columnwidth]{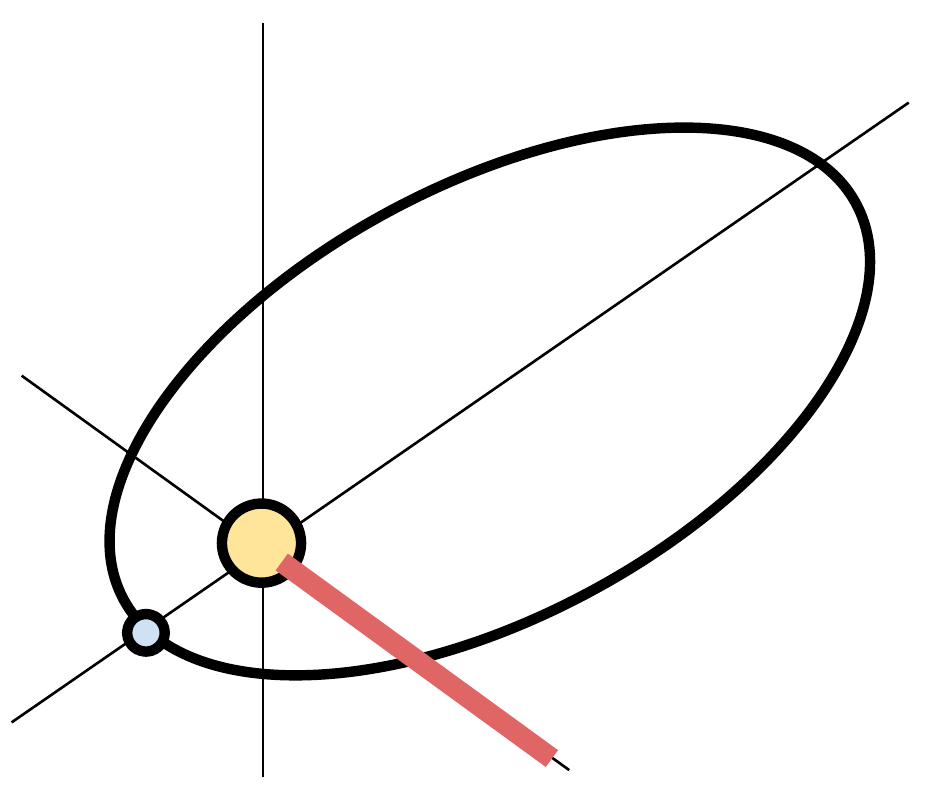} & \includegraphics[width=0.18\columnwidth]{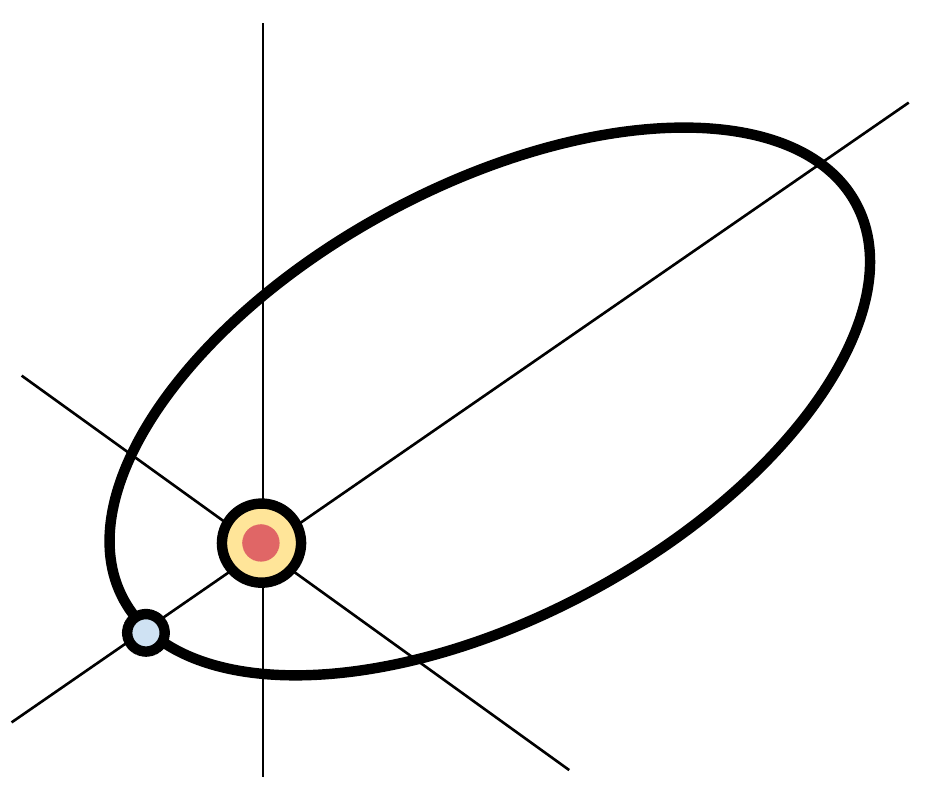} & \includegraphics[width=0.18\columnwidth]{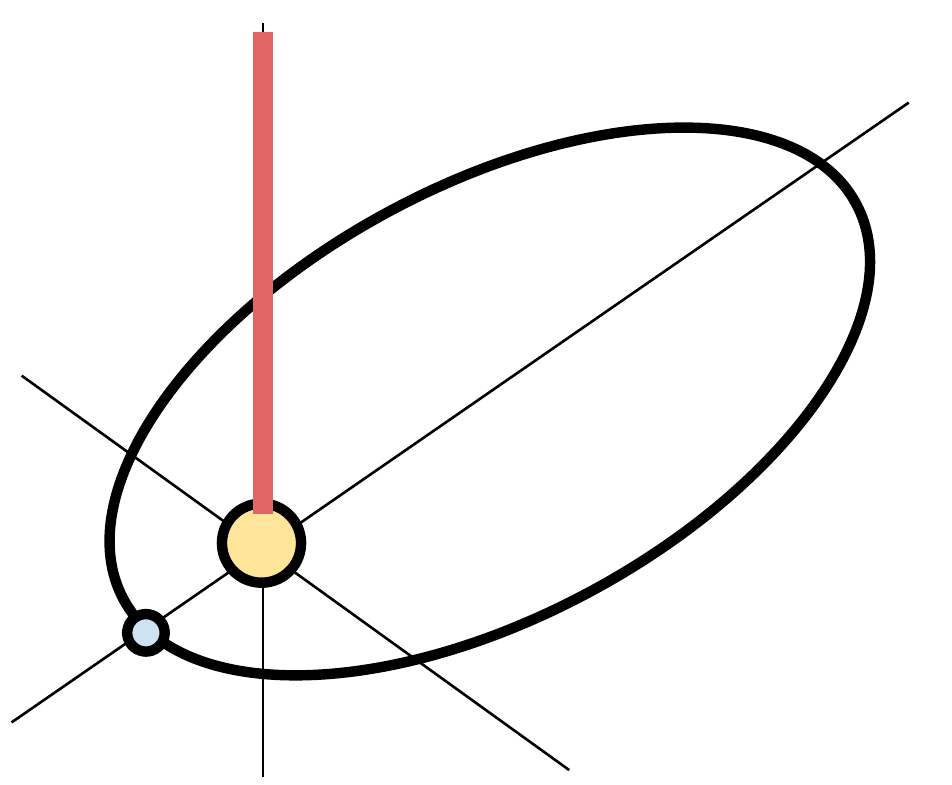}
\end{tabular}
\caption{The viewing angles used in Figure~\ref{twoFig}. The sketches show the orbit of an eccentric system, with the planet at periapse (which we define as phase $= 0$). Each line-of-sight is shown as a thick red line. For the oblique case the viewing angle projection is directly out of the page.}
\label{viewingAngles}
\end{table}

\subsection{The dinosaur in the detail}

\begin{figure*}
\centering
\includegraphics[width=\textwidth]{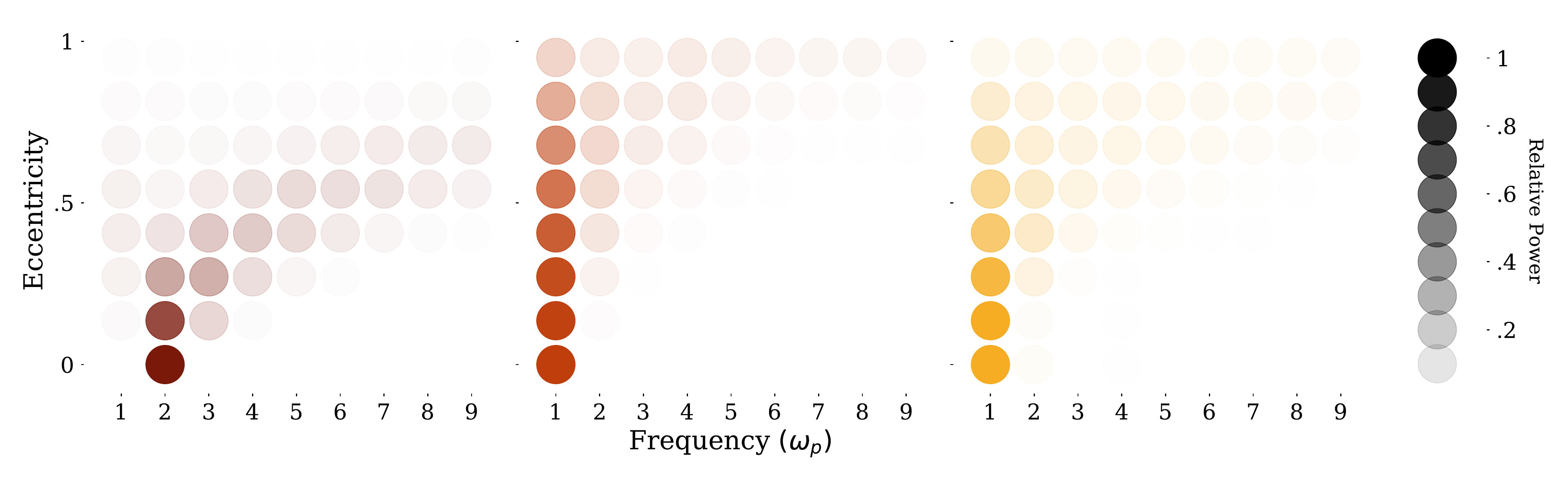}
\caption{The eccentricity dependence of individual harmonics in the power spectrum, shown individually for tides (dark red), beaming (brick red) and reflections (yellow). The same system parameters are used as in Figure~\ref{intro} save for varying eccentricity. The relative power has been normalised independently for each effect (colour) and each eccentricity (row).
}
\label{eccentricity}
\end{figure*}

Figure~\ref{eccentricity} represents visually the relative strength of power spectrum peaks at the first few harmonics as a function of eccentricity.

In this figure, we can see clearly that tidal signatures peak at higher and higher harmonics as eccentricity increases (yielding those characteristic ``stegosaurus spines"). For $e>0.5$, the spectrum is spread over so many peaks as to make the relative power almost invisible in these plots, though the total power summed over all harmonics increases rapidly for increasing eccentricity.

The beaming and reflection signatures have a strong peak at the $1^{st}$ harmonic, with more power in higher harmonics for larger $e$. At any chosen eccentricity, the signal from reflection has significantly more power in higher harmonics than the beaming signal.

Let us take a moment here to build some intuition as to where these higher-order harmonics originate from.

As detailed in section~\ref{photometric}, simple intuition can be applied to the circular case. Let us apply similarly simple intuition to the near-radial case ($e \rightarrow 1$). As the eccentricity approaches unity, any signal will occur over a shorter time window as the time the planet spends near periapse decreases. Thus, in the limit of extreme eccentricities, any signal can be approximated as a $\delta$ function, occurring once per orbit. The fourier transform of a $\delta$ function has constant amplitude across all frequency. Thus, in the limit of high eccentricities, the power spectra of each of these signals should tend to having nearly equal power extending up to high harmonics. Indeed this is what we see in Figure~\ref{eccentricity}.

For small eccentricities, we could explicitly derive the amplitude of specific peaks, and the tools needed to do this are mapped out in Appendix \ref{analytic}. Summarising some of that logic here, for small eccentricities we can expand various terms in powers of $e \cos \eta$ and $e \sin \eta$ (and also can translate functions of $\Phi$ to functions of $\eta$). Thus, to first order in $e$, the signal depends on higher powers of $\cos \eta$ and $\sin \eta$ and thus a Fourier transform of the signal has power at higher harmonics.

We derive (end of Appendix \ref{analytic}) the amplitude of the third harmonic in the specific case of a tidal dominated signal seen edge-on, with periapse aligned with the viewer. We find the third (first) harmonic has an amplitude a factor of $\frac{7e}{2}$ ($\frac{e}{2}$) of the peak at the second harmonic.


\section{Prospects for detection}
\label{prospects}

We now move to discussion of the prospects and potential difficulties of the detection these higher-order harmonics in real stellar light curves. In particular, we will focus on the Transiting Exoplanet Survey Satellite (TESS, \citealt{Ricker15}).

\subsection{Power spectra from TESS}
TESS is an all-sky survey which measures the photometric variation of 200,000 target stars at 2-minute cadence, as well as takes full-frame images of its entire $2300\,\mathrm{deg}^2$ field-of-view at 30-minute cadence. Over its two-year primary mission, it will observe azimuthal slices of first the southern and then the northern ecliptic hemisphere, spending $\sim 27$ days on each slice. Regions of the sky that belong to more than one overlapping slice benefit from a longer observational baseline, with small ``continuous-viewing" zones of $\sim 900\,\mathrm{deg}^2$ near the ecliptic poles observed for $\sim 351$ days each.

TESS is expected to achieve a photometric precision of roughly 50 parts per million (and indeed, the cleanest target star observations are already meeting this goal as of the Sector 1 data release\footnote{\url{https://heasarc.gsfc.nasa.gov/docs/tess/observing-technical.html}, accessed 1 April 2019.}). If the period of an observed planetary system can be derived, the data can be folded over that period and binned, reducing the photometric uncertainty by a factor of $\sqrt{N}$, where $N$ is the number of observed periods. Thus, planets with small semi-major axes, and correspondingly short periods, will be the most promising for detection and characterisation.

\begin{figure*}
\centering
\includegraphics[width=\textwidth]{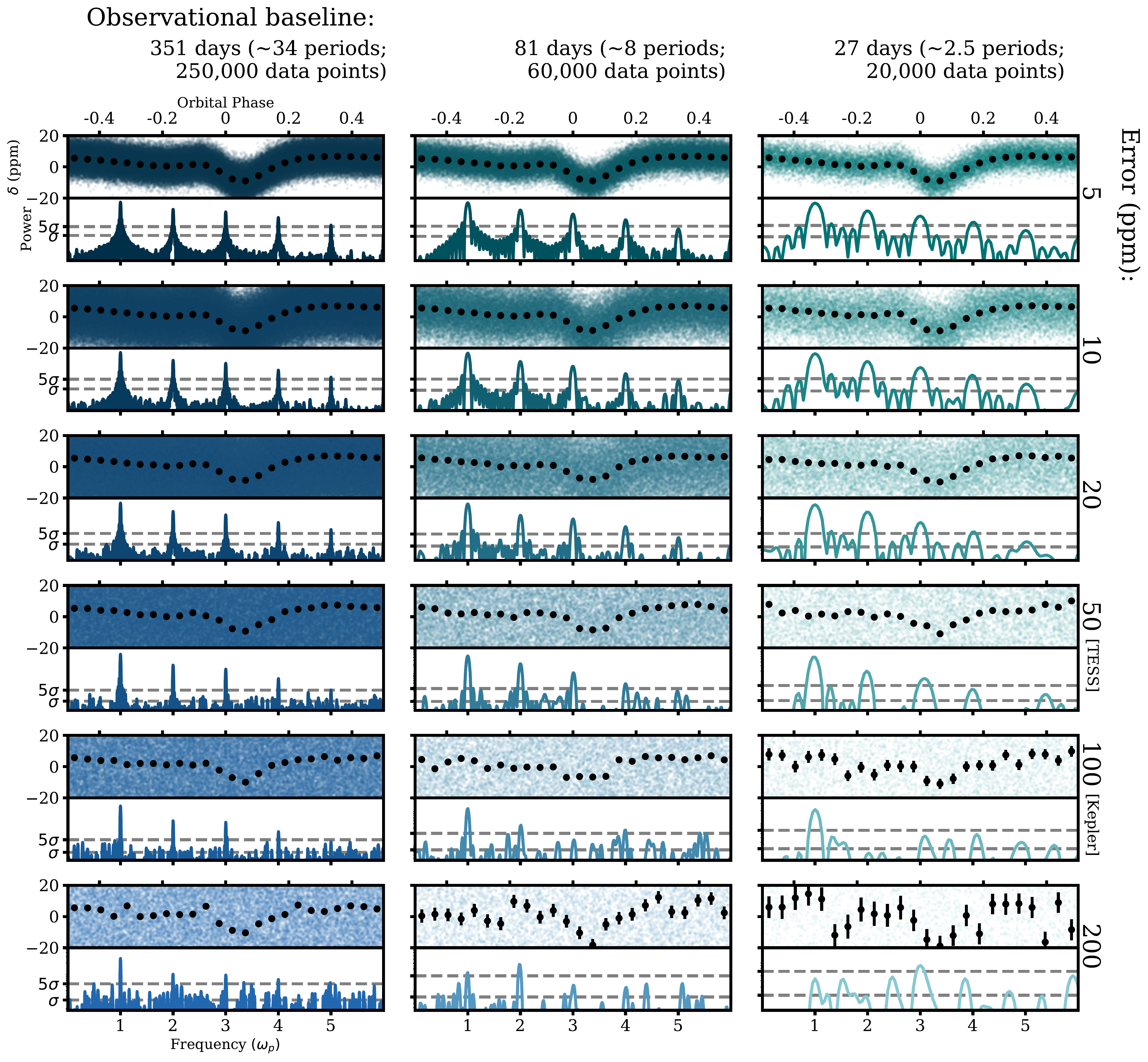}
\caption{Mock light curves and power spectra for the planetary system of Figure~\ref{intro}, calculated with varying levels of Gaussian-distributed noise over varying observational baselines. Descending rows have higher photometric uncertainty, and columns, moving to the right, have shorter observational baselines. The uncertainty and time sampling is constructed to be directly comparable to the TESS survey (more details in text). Individual data points in the light curve, folded over one period, are shown in colour, and the mean values (averaged over 6-hour bins) are shown in black. The mean and standard deviation of the power spectra are calculated, and the plots show the mean (coinciding with the x-axis) and 1 and 5 standard deviations above this (on a log scale) as dashed gray horizontal lines.}
\label{error}
\end{figure*}

Figure~\ref{error} shows mock light curves and power spectra for our example planet, given a range of photometric uncertainties and observational baselines. The light curves are sampled at TESS's 2-minute cadence, with baselines corresponding to the 351, 81, and 27 day-baseline viewing regions of TESS. The simulated uncertainties are Gaussian, using the stated uncertainty in each row of subplots as the width of the distribution. 

There is no red noise included in our calculations, which one might expect from stellar activity. In general such noise is incoherent when the light curve is phase-folded at the planet's orbital period, so we do not expect it to contribute meaningful uncertainty in the case of short-period planets with many observed transits. (See \citealt{Shporer17} for a detailed discussion of the sources of stellar noise and their impact on the observable precision.)

The 50 parts per million uncertainty row is most relevant to a TESS observation of a system like our example planet. We see that for all but the shortest TESS observational baseline, the second and third harmonics in the power spectrum are clearly visible, though their relative amplitudes can vary.

Higher photometric precision is obtainable with instruments such as the Hubble Space Telescope \citep{Demory2015} and the forthcoming James Webb Space Telescope \citep{Beichman14}, for which errors of order of 10-20 ppm per period may be attainable. For instruments such as these, an observation of our example system over two periods might be sufficient to resolve the $3^{\mathrm{rd}}$ harmonic and above. The photometric precision of the \textit{Kepler} mission is of order 100 ppm; higher-order harmonics of this example planet would be perfectly observable at this precision over \textit{Kepler}'s four-year observational baseline at \textit{Kepler}'s short observational cadence ($\sim 1$ minute), although this is not shown in Figure~\ref{error}. As discussed in PS18, many other confirmed exoplanets should also have detectable out-of-transit signals in archival \textit{Kepler} data.

Our example planet has a characteristic $\delta$ of order 10 ppm, but many systems exist for which a significantly larger-amplitude signal would be expected. For larger characteristic $\delta$, the obtainable signal-to-noise ratio increases; for intuition, if the characteristic $\delta$ were of order 100 ppm (instead of our example planet's 10 ppm), the power spectrum plotted in Figure~\ref{error} for the 5 ppm error is more representative, in terms of signal-to-noise, of the expected TESS observation. Similarly, shorter-period planets will be much better constrained over the same observational baseline, as more periods can be overlaid.

\subsection{Yield from TESS}

We can also make some rudimentary estimations of the number of planetary systems with detectable OoT effects (``OoT systems") we will observe in the TESS survey, and how many of these will have observable higher order harmonics.

\citet{Barclay18} simulated the yield of TESS planet discoveries we can expect based on our current understanding of exoplanet occurrence rates, multiplicities, and properties (including planet periods, radii, and eccentricities). They selected stars from the TESS Input Catalog \citep{Stassun18}, which includes stellar mass and radius, cumulative observation time over the TESS two-year primary mission (based on the star's on-sky position), and estimates of stellar noise over a one hour integration for each member, and estimated which stars would be good candidates as postage-stamp targets (observed at 2-minute cadence) vs. stars observed only in the full-frame images (FFIs), at 30-minute cadence. Given this chosen catalog of target and FFI stars, they then simulated planets around these stars and asked which planets would be detected to transit at least twice at SNR $\geq 7.3$.

Their reported catalog includes only planets they categorise as ``detectable" according to the above criteria, thus discarding the vast majority which do not transit, as well as any with a long period or small transit depth compared to the level of photometric noise from stellar activity. They predict a little over 4,000 detectable systems, a number that can vary by as much as a factor of two depending on the assumptions and demands made of the data.

Here we use their catalog of ``detected" simulated planets to estimate the yield of OoT systems we might expect from TESS. Firstly, in order to calculate the amplitude of the tidal and beaming effects in any particular system, we also need to estimate the mass of the planet, which we do via the mass-radius model \texttt{forecaster} \citep{Chen17}. We then set the albedo of all planets to $A_g = 0.15$, the polar angle to $\theta_v =\frac{\pi}{2}$ (because these planets transit, and therefore must be observed nearly edge-on), and choose azimuthal angles from a uniform distribution over the range $\phi_v \sim [0,2\pi]$.

We calculate the maximum variability in flux due to OoT effects, $\delta_{OoT}$, (using the relationships in Section \ref{photometric}) for every planet in the sample, as well as an estimate for the amplitude of the third harmonic (see Appendix \ref{tidesCorrection} for details). We also calculate an expected noise threshold for each system as the inherent stellar noise over a one-hour integration (given in the TESS input catalog), divide by the square root of the number of observed planet periods (which depends both on the planet's period and the star's position in the TESS observing sectors). For each planet, we calculate a signal-to-noise ratio as $\delta_{OoT}$ divided by this noise threshold.

\begin{figure*}
\includegraphics[width=\textwidth]{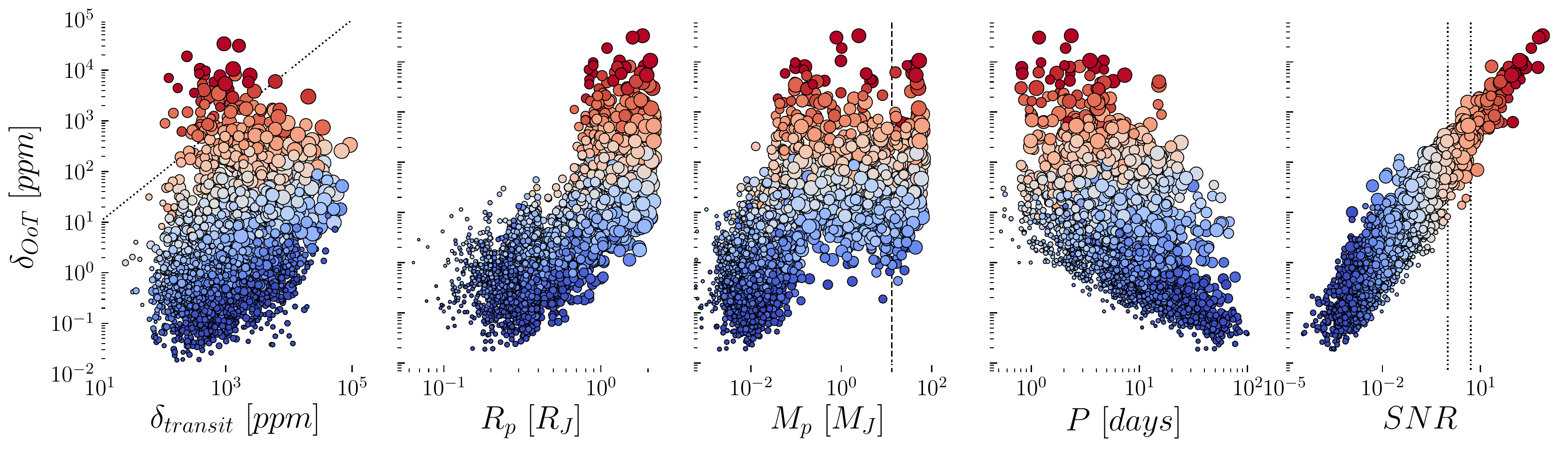}
\caption{An estimation of the magnitude of the out-of-transit (OoT) signal $\delta_{OoT}$ across the predicted yield of transiting TESS planets from \citet{Barclay18}. Each point is a predicted planet, size-scaled by its planetary radius and colour-coded by the signal-to-noise ratio of the OoT effects. From left to right, we compare $\delta_{OoT}$ to: estimated transit depth, planetary radius, planetary mass, planetary period, and the signal to noise ratio of the OoT effects. In the first subplot, the dotted line shows the boundary upon which the magnitude of OoT signals is equal to the transit depth (i.e., the magnitude of the transit signal). In the third subplot, the dashed line shows the mass above which we would consider the object a brown dwarf ($\sim 13 M_J$). In the final plot the vertical dotted lines mark signal-to-noise ratios of 1 and 5. Of the predicted 4373 TESS transiting planets, 262 have an OoT signal greater than their noise threshold (of which 60 might be considered brown dwarfs), and 102 have $SNR>5$.}
\label{yield}
\end{figure*}

In Figure \ref{yield}, we plot our calculated $\delta_{OoT}$ for the planets simulated by \cite{Barclay18}. We see that the majority of the systems with SNR$\geq1$, i.e. OoT variability greater than the noise threshold, are hot Jupiters or similar. Out of a sample of 4373, we predict 262 systems with SNR $\geq1$ (and, out of these, 106 systems with observable phase curve amplitude at the third harmonic). For a more conservative estimate: 102 of these 262 OoT systems are detectable at SNR $\geq 5$.

Examining the distribution of OoT signals with respect to the parameters of the system, we see that the highest-SNR OoT planets are likely to close in size to Jupiter, with masses of Neptune or above, and short periods.

By using this data set, we have limited ourselves to only transiting planets. These are much easier to detect and constrain and so provide an excellent first sample of OoT effects visible in TESS. However, many systems that might have observable OoT signals will not be transiting - the probability of observing a transit goes as $\frac{R}{a}$, whilst OoT signals are always present, though their amplitude drops with $\sin \theta_v$. We can make a very rough estimate of the total number of observable OoT signals in TESS, using the fact that for every observed transit we can expect roughly $\frac{a}{R}$ similar non-transiting systems to exist. 

By this logic, we estimate that the total number of TESS OoT systems with SNR $\geq 1$ is 1221 (327 with SNR $\geq 5$).

There are a number of factors that make this estimate very approximate, including:

\begin{itemize}
\item Some of the systems predicted here may be considered brown dwarfs, as \texttt{forecaster} maps some fraction of Jupiter sized objects onto masses above the deuterium burning limit. 60 of our estimated 262 planets with $SNR > 1$ have masses $> 13 M_J$ (see Figure~\ref{yield}, panel 3).
\item We are extrapolating only from systems which transit twice or more---moreover, the \citep{Barclay18} simulated yield excludes grazing transits, a nuance that our correction for transit probability does not capture.
\item In calculating the noise threshold for each star, We have not specified the form of the noise---it may be white noise, occasional violent events (like stellar flares), or (quasi-)periodic stellar variability, such as star spots or pulsations. In the case of the latter, if the stellar rotation period is distinct from the planet's orbital period, the stellar-periodic noise can be accurately removed numerically. However, if the stellar rotation period is very close to the planet's orbital period, it can become impossible to distinguish.
\item The \cite{Barclay18} synthetic population of planets is based on occurrence rates and planet properties extrapolated from currently-known exoplanets. If the true underlying distribution of exoplanets is poorly sampled or contains features we have not yet discerned, then this may be an unrepresentative sample---particularly because TESS target stars are, generally speaking, cooler than \textit{Kepler} target stars.
\item Here we have taken most of the planet and star properties to be uncorrelated, but for example if the stellar noise is strongly correlated with planetary properties, we may find that OoT signals are easier or harder to detect across different regions of planet parameter space.
\item We have, for our noise threshold calculation, used the reported noise over a one-hour integration of the stellar flux. However, as OoT effects are visible over the whole planet period, it is possible to bin light curve measurements to much longer effective integration times in search of the OoT signal. Binning could reduce the stellar noise by a factor of 2 or more for some stars.
\end{itemize}

As detections of exoplanets by TESS start to mount up, it will be interesting to compare our rough prediction with the evolving population.

\subsection{Dependence on system properties}

So far, we have focused only on a small slice of the possible parameter space of planetary systems. In PS18, we showed the relative strength of photometric (and spectroscopic) effects for already-confirmed exoplanets, but here we explore the parameter space of possible planetary systems in the abstract.

\begin{figure}
\includegraphics[width=\columnwidth]{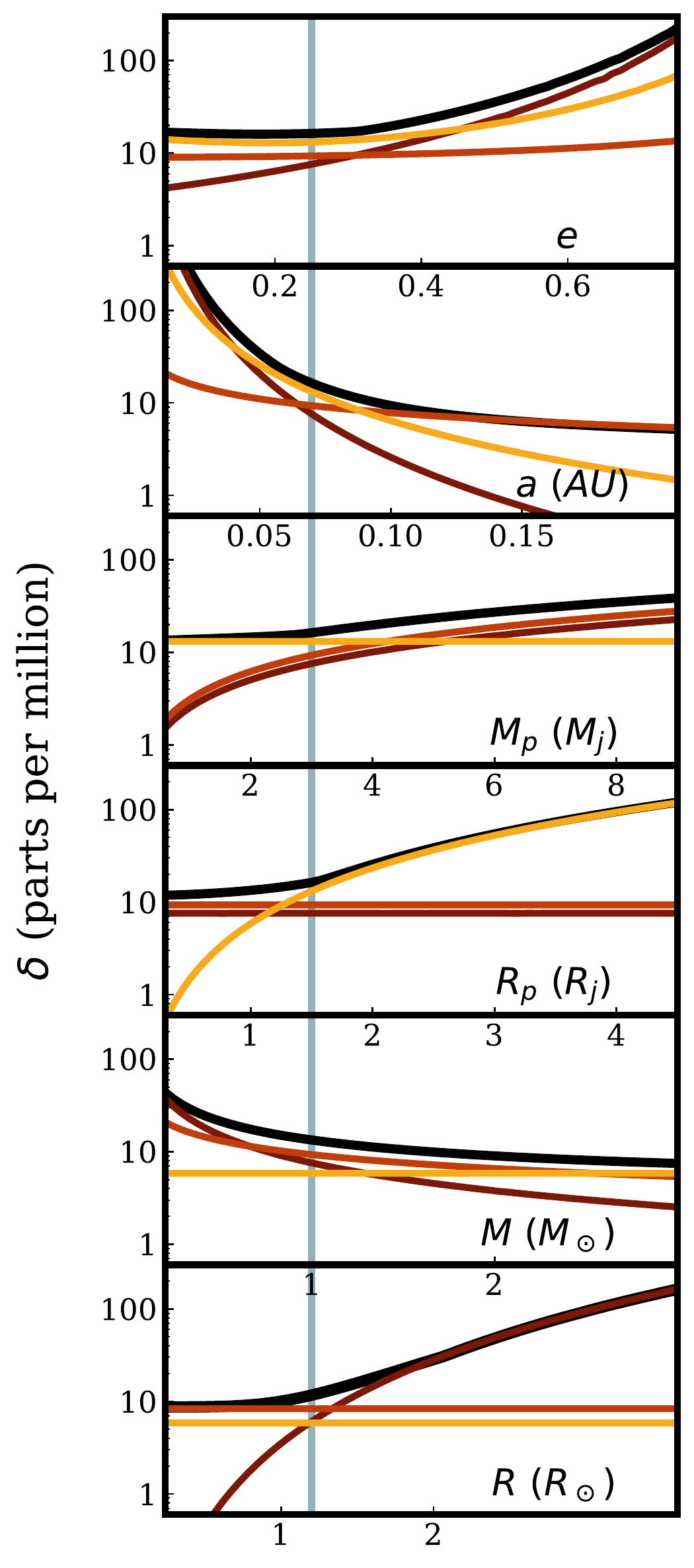}
\caption{The maximum variation in the light curve due to tides (dark red), beaming (brick red) and reflections (yellow), compared to the total variation (black). The grey line shows the properties of our example planet (as used in Figure \ref{intro} and throughout) and the properties of the system are varied independently from these values. All x-scaling is linear and the y-scale is the same for all panels.}
\label{compare}
\end{figure}

Figure~\ref{compare} shows how the maximum $\delta_t,\ \delta_r,\ \delta_b,$ and $\delta_\Sigma$ over one planetary orbit differs for various systems.

In each plot, the parameters of our example planet (see Figure \ref{intro}) are used, save for the one which we vary. This is more physically realistic for some parameters than others: for example, two equivalent planets with very different semi-major axes can exist, but we might be more surprised by two planets with the same mass and yet markedly different radii. As such, we vary the parameters only over a small (linear) range. 

We also only use one projection, the same as in Figure \ref{intro}, though the results will vary a little when the system is viewed from other angles.

We will go through each panel in the order shown, highlighting which signal, if any, dominates in a particular parameter space:
\begin{itemize}
\item Highly eccentric planets - For a fixed semi-major axis, more eccentric systems have much larger tidal signatures, as tides have the strongest dependence on pericenter distance.
\item Large orbits - Beaming is dominant for larger orbits (though the amplitude of the signal is small), whilst tides dominate for orbits passing close to the host star.
\item Massive planets - Both tides and beaming have the same dependence on planetary mass, with larger $\delta$ for heavier planets. Reflection signals will dominate for lower masses (this remains true under the assumption of constant planetary density, with $M_p \propto R_p^3$).
\item Giant planets - Only the reflection signal depends on planetary radius, and thus dominates for larger planets (of fixed mass).
\item Massive stars - The reflection signal strength is independent of stellar mass, though if we assume constant stellar density (with $M \propto R^3$) the tidal signature is also constant and may dominate. At lower masses, for fixed radii, tides dominate as the outer layers of the star are less gravitationally bound and tidal distortions are larger.
\item Giant stars - For larger radii, tides dominate, as the distortion of the outer layers becomes larger.
\end{itemize}

Alternatively, we could ask which regimes of parameter space each signal dominates in:
\begin{itemize}
\item Tides - Tidal signatures dominate when the outer layers of the star are easily distorted. This occurs for close orbits and massive planets (where the pull of the planet is larger) and for large, low-mass stars (where the competing gravity of the star is reduced). Thus tides are promising for examining eccentric or close-orbiting hot Jupiters, brown dwarfs, or planets around giant stars. The signals can be very large (up to percentage-level changes in luminosity) and in eccentric systems are visible from all angles.
\item Beaming - Beaming signals are not very large (at most $\delta_b \sim 10$ ppm for the parameters of these systems), but they are the least dependent on the distance between the planet and the star. Thus they may be of interest in the largest number of systems, though only with sufficient photometric precision. Currently they are of most interest for examining massive planets on large orbits (where ``large" here still only refers to orbits closer to their star than 1 AU) and planets which are already constrained by existing radial velocity measurements. 
\item Reflections - The fraction of a star's light reflected by a planet is strongly dependent on the planetary radius, and has no direct dependence on the mass of the planet or the star (though assuming planets have uniform density, $\delta_l \propto M_p^{\frac{2}{3}}$) and thus is of most interest in low-mass or large-radius planets and planets around high-mass stars. Like tides, the signals can be large (up to parts per thousand levels) and for eccentric systems they can be visible from all viewing angles.
\end{itemize}

\section{Discussion and conclusions}\label{discussion}

In this paper, we have shown that the power spectra of light curves of stars hosting eccentric planets, even at small orbital eccentricities, contain higher-order harmonics that cannot be explained by the assumption of circular orbits.

Variations in the star's apparent luminosity due to tides, beaming, and/or reflections all give power spectra with peaks at 3 times the planet's orbital frequency and above. If tides dominate, the spectrum can extend to very high harmonics, and the dominant frequency may itself be a high harmonic.

In Section \ref{prospects}, we show that these effects will be visible in surveys such as TESS and \textit{Kepler} for many known planets, and that more will be detectable in both existing and future data. As shown in PS18, eccentric planets should be visible regardless of viewing angle for systems with strong reflection or tidal signals.
 
We also discuss the strength of these three signals across the possible parameter space for systems hosting exoplanets. We show that tides will give the largest possible signals, especially for eccentric hot-Jupiters and giant stars. Reflections will dominate for lower-mass planets, whilst beaming is of most interest for planets on larger orbits (though the signal will be small).

We provide open-source code, \texttt{OoT}, for calculating the photometric effect of tides, beaming, and reflections as a function of time and the system properties. This is detailed in Appendix \ref{code}.

\subsection{Using power spectra as a tool for planet detection}

Here, we have highlighted observable effects in the power spectrum due to the photometric signals of tides, beaming, and reflection, each of which depends differently upon the parameters of the planetary system in question. Consequently, we might hope to use power spectra to characterise planetary systems. However, this is considerably beyond the scope of the present work.

Simple diagnostics are possible when the power spectrum is both well-resolved and dominated by a single signal of the three, or when the parameters of the system are very well constrained by independent observations.

However, in most systems, there will be significant degeneracies between the contributions to the power spectrum of different out-of-transit signals. Measurement error will further complicate any attempt to fit data to predicted power spectra. In short, power spectra are not currently an independently useful tool for characterising planets. It will be fitting models to the light curve, not the power spectrum, that will yield tight constraints on system properties.

In spite of the difficulty of using power spectra to characterise planets, power spectra are by far the most useful and easily accessible tool for \textit{finding} planets (or at least hinting at their existence). Power spectra are relatively easy to produce for any light curve, as they require no foreknowledge of the system, and preserve only the periodic photometric signals present in the light curve.

Most out-of-transit signals are low-amplitude, and likely very difficult to observe in the light curve over a single period. It is only when data are stacked over many planetary periods that the signal becomes clear and models can be fitted. But this requires knowledge of the presence of a planet, and of its period. This is the invaluable information that the power spectrum tells us.

Where once planets were detected through a single clear signal, such as a transit, we are now entering an era where signs of the same system can be observed via many different methods. The out-of-transit effects discussed here are small, but they are universal. They depend heavily on multiple parameters of the system, which makes them complex to model, but also breaks degeneracies and tells us about myriad aspects of the planet, star, and orbit. 

Particularly as we seek to stretch not just the number, but the range of known planets and their properties, we will have to leverage small-amplitude signals and the combined effect of many independent observations and physical processes. The power spectrum is simply one, though will often be the first, of the tools needed to do this.

\section*{Acknowledgements}

We thank Cathie Clarke, Nicholas C. Stone, Adam Wheeler, Alex Teachey, Tiffany Jansen, David Kipping and the anonymous reviewer for their helpful comments and discussions. Z.P. acknowledges support from the UK Science and Technologies Facilities Council (STFC).

\bibliographystyle{mnras}
\bibliography{bib}

\begin{thebibliography}{}
\makeatletter
\relax
\def\mn@urlcharsother{\let\do\@makeother \do\$\do\&\do\#\do\^\do\_\do\%\do\~}
\def\mn@doi{\begingroup\mn@urlcharsother \@ifnextchar [ {\mn@doi@}
  {\mn@doi@[]}}
\def\mn@doi@[#1]#2{\def\@tempa{#1}\ifx\@tempa\@empty \href
  {http://dx.doi.org/#2} {doi:#2}\else \href {http://dx.doi.org/#2} {#1}\fi
  \endgroup}
\def\mn@eprint#1#2{\mn@eprint@#1:#2::\@nil}
\def\mn@eprint@arXiv#1{\href {http://arxiv.org/abs/#1} {{\tt arXiv:#1}}}
\def\mn@eprint@dblp#1{\href {http://dblp.uni-trier.de/rec/bibtex/#1.xml}
  {dblp:#1}}
\def\mn@eprint@#1:#2:#3:#4\@nil{\def\@tempa {#1}\def\@tempb {#2}\def\@tempc
  {#3}\ifx \@tempc \@empty \let \@tempc \@tempb \let \@tempb \@tempa \fi \ifx
  \@tempb \@empty \def\@tempb {arXiv}\fi \@ifundefined
  {mn@eprint@\@tempb}{\@tempb:\@tempc}{\expandafter \expandafter \csname
  mn@eprint@\@tempb\endcsname \expandafter{\@tempc}}}

\bibitem[\protect\citeauthoryear{{Akeson} et~al.,}{{Akeson}
  et~al.}{2013}]{Akeson13}
{Akeson} R.~L.,  et~al., 2013, \mn@doi [Publications of the Astronomical
  Society of the Pacific] {10.1086/672273}, \href
  {https://ui.adsabs.harvard.edu/\#abs/2013PASP..125..989A} {125, 989}

\bibitem[\protect\citeauthoryear{{Armstrong} \& {Rein}}{{Armstrong} \&
  {Rein}}{2015}]{Armstrong15}
{Armstrong} C.,  {Rein} H.,  2015, \mn@doi [\mnras] {10.1093/mnrasl/slv112},
  \href {http://adsabs.harvard.edu/abs/2015MNRAS.453L..98A} {453, L98}

\bibitem[\protect\citeauthoryear{{Barclay}, {Pepper}  \& {Quintana}}{{Barclay}
  et~al.}{2018}]{Barclay18}
{Barclay} T.,  {Pepper} J.,   {Quintana} E.~V.,  2018, \mn@doi [\apjs]
  {10.3847/1538-4365/aae3e9}, \href
  {https://ui.adsabs.harvard.edu/abs/2018ApJS..239....2B} {239, 2}

\bibitem[\protect\citeauthoryear{{Beichman} et~al.,}{{Beichman}
  et~al.}{2014}]{Beichman14}
{Beichman} C.,  et~al., 2014, \mn@doi [\pasp] {10.1086/679566}, \href
  {https://ui.adsabs.harvard.edu/abs/2014PASP..126.1134B} {126, 1134}

\bibitem[\protect\citeauthoryear{{Bento} et~al.,}{{Bento}
  et~al.}{2018}]{Bento18}
{Bento} J.,  et~al., 2018, \mn@doi [\mnras] {10.1093/mnras/sty726}, \href
  {http://adsabs.harvard.edu/abs/2018MNRAS.477.3406B} {477, 3406}

\bibitem[\protect\citeauthoryear{{Binney} \& {Tremaine}}{{Binney} \&
  {Tremaine}}{2008}]{Binney08}
{Binney} J.,  {Tremaine} S.,  2008, {Galactic Dynamics: Second Edition}.
Princeton University Press

\bibitem[\protect\citeauthoryear{{Bonomo} et~al.,}{{Bonomo}
  et~al.}{2015}]{Bonomo15}
{Bonomo} A.~S.,  et~al., 2015, \mn@doi [\aap] {10.1051/0004-6361/201323042},
  \href {http://adsabs.harvard.edu/abs/2015A%26A...575A..85B} {575, A85}

\bibitem[\protect\citeauthoryear{{Charbonneau}, {Noyes}, {Korzennik},
  {Nisenson}, {Jha}, {Vogt}  \& {Kibrick}}{{Charbonneau}
  et~al.}{1999}]{Charbonneau99}
{Charbonneau} D.,  {Noyes} R.~W.,  {Korzennik} S.~G.,  {Nisenson} P.,  {Jha}
  S.,  {Vogt} S.~S.,   {Kibrick} R.~I.,  1999, \mn@doi [\apjl]
  {10.1086/312234}, \href {http://adsabs.harvard.edu/abs/1999ApJ...522L.145C}
  {522, L145}

\bibitem[\protect\citeauthoryear{{Charbonneau}, {Brown}, {Latham}  \&
  {Mayor}}{{Charbonneau} et~al.}{2000}]{Charbonneau00}
{Charbonneau} D.,  {Brown} T.~M.,  {Latham} D.~W.,   {Mayor} M.,  2000, \mn@doi
  [\apjl] {10.1086/312457}, \href
  {http://adsabs.harvard.edu/abs/2000ApJ...529L..45C} {529, L45}

\bibitem[\protect\citeauthoryear{{Chen} \& {Kipping}}{{Chen} \&
  {Kipping}}{2017}]{Chen17}
{Chen} J.,  {Kipping} D.,  2017, \mn@doi [\apj] {10.3847/1538-4357/834/1/17},
  \href {https://ui.adsabs.harvard.edu/abs/2017ApJ...834...17C} {834, 17}

\bibitem[\protect\citeauthoryear{{Cowan} \& {Agol}}{{Cowan} \&
  {Agol}}{2011}]{Cowan11}
{Cowan} N.~B.,  {Agol} E.,  2011, \mn@doi [\apj] {10.1088/0004-637X/729/1/54},
  \href {http://adsabs.harvard.edu/abs/2011ApJ...729...54C} {729, 54}

\bibitem[\protect\citeauthoryear{{Cowan}, {Chayes}, {Bouffard}, {Meynig}  \&
  {Haggard}}{{Cowan} et~al.}{2017}]{Cowan17}
{Cowan} N.~B.,  {Chayes} V.,  {Bouffard} {\'E}.,  {Meynig} M.,   {Haggard}
  H.~M.,  2017, \mn@doi [\mnras] {10.1093/mnras/stx133}, \href
  {http://adsabs.harvard.edu/abs/2017MNRAS.467..747C} {467, 747}

\bibitem[\protect\citeauthoryear{{Demory} et~al.,}{{Demory}
  et~al.}{2015}]{Demory2015}
{Demory} B.-O.,  et~al., 2015, \mn@doi [\mnras] {10.1093/mnras/stv673}, \href
  {http://adsabs.harvard.edu/abs/2015MNRAS.450.2043D} {450, 2043}

\bibitem[\protect\citeauthoryear{{Esteves}, {De Mooij}  \&
  {Jayawardhana}}{{Esteves} et~al.}{2013}]{Esteves13}
{Esteves} L.~J.,  {De Mooij} E.~J.~W.,   {Jayawardhana} R.,  2013, \mn@doi
  [\apj] {10.1088/0004-637X/772/1/51}, \href
  {http://adsabs.harvard.edu/abs/2013ApJ...772...51E} {772, 51}

\bibitem[\protect\citeauthoryear{{Faigler} \& {Mazeh}}{{Faigler} \&
  {Mazeh}}{2011}]{Faigler11}
{Faigler} S.,  {Mazeh} T.,  2011, \mn@doi [\mnras]
  {10.1111/j.1365-2966.2011.19011.x}, \href
  {http://adsabs.harvard.edu/abs/2011MNRAS.415.3921F} {415, 3921}

\bibitem[\protect\citeauthoryear{{Fuller}}{{Fuller}}{2017}]{Fuller17}
{Fuller} J.,  2017, \mn@doi [\mnras] {10.1093/mnras/stx2135}, \href
  {http://adsabs.harvard.edu/abs/2017MNRAS.472.1538F} {472, 1538}

\bibitem[\protect\citeauthoryear{{Gai} \& {Knuth}}{{Gai} \&
  {Knuth}}{2018}]{Gai18}
{Gai} A.~D.,  {Knuth} K.~H.,  2018, \mn@doi [\apj] {10.3847/1538-4357/aa9ee1},
  \href {https://ui.adsabs.harvard.edu/#abs/2018ApJ...853...49G} {853}

\bibitem[\protect\citeauthoryear{{Jackson}, {Lewis}, {Barnes}, {Drake Deming},
  {Showman}  \& {Fortney}}{{Jackson} et~al.}{2012}]{Jackson12}
{Jackson} B.~K.,  {Lewis} N.~K.,  {Barnes} J.~W.,  {Drake Deming} L.,
  {Showman} A.~P.,   {Fortney} J.~J.,  2012, \mn@doi [\apj]
  {10.1088/0004-637X/751/2/112}, \href
  {http://adsabs.harvard.edu/abs/2012ApJ...751..112J} {751, 112}

\bibitem[\protect\citeauthoryear{{Jansen} \& {Kipping}}{{Jansen} \&
  {Kipping}}{2018}]{Jansen17}
{Jansen} T.,  {Kipping} D.,  2018, \mn@doi [\mnras] {10.1093/mnras/sty1149},
  \href {https://ui.adsabs.harvard.edu/abs/2018MNRAS.478.3025J} {478, 3025}

\bibitem[\protect\citeauthoryear{{Kane} \& {Gelino}}{{Kane} \&
  {Gelino}}{2012}]{Kane12}
{Kane} S.~R.,  {Gelino} D.~M.,  2012, \mn@doi [\mnras]
  {10.1111/j.1365-2966.2012.21265.x}, \href
  {http://adsabs.harvard.edu/abs/2012MNRAS.424..779K} {424, 779}

\bibitem[\protect\citeauthoryear{{Kane}, {Ciardi}, {Gelino}  \& {von
  Braun}}{{Kane} et~al.}{2012}]{Kane12b}
{Kane} S.~R.,  {Ciardi} D.~R.,  {Gelino} D.~M.,   {von Braun} K.,  2012,
  \mn@doi [\mnras] {10.1111/j.1365-2966.2012.21627.x}, \href
  {http://adsabs.harvard.edu/abs/2012MNRAS.425..757K} {425, 757}

\bibitem[\protect\citeauthoryear{{Kopal}}{{Kopal}}{1959}]{Kopal59}
{Kopal} Z.,  1959, {Close binary systems}

\bibitem[\protect\citeauthoryear{{Kreidberg}}{{Kreidberg}}{2015}]{Kreidberg15}
{Kreidberg} L.,  2015, \mn@doi [\pasp] {10.1086/683602}, \href
  {http://adsabs.harvard.edu/abs/2015PASP..127.1161K} {127, 1161}

\bibitem[\protect\citeauthoryear{{Kumar}, {Ao}  \& {Quataert}}{{Kumar}
  et~al.}{1995}]{Kumar95}
{Kumar} P.,  {Ao} C.~O.,   {Quataert} E.~J.,  1995, \mn@doi [\apj]
  {10.1086/176055}, \href {http://adsabs.harvard.edu/abs/1995ApJ...449..294K}
  {449, 294}

\bibitem[\protect\citeauthoryear{{Lillo-Box} et~al.,}{{Lillo-Box}
  et~al.}{2014}]{LilloBox14}
{Lillo-Box} J.,  et~al., 2014, \mn@doi [\aap] {10.1051/0004-6361/201322001},
  \href {http://adsabs.harvard.edu/abs/2014A%26A...562A.109L} {562, A109}

\bibitem[\protect\citeauthoryear{{Loeb} \& {Gaudi}}{{Loeb} \&
  {Gaudi}}{2003}]{Loeb03}
{Loeb} A.,  {Gaudi} B.~S.,  2003, \mn@doi [\apjl] {10.1086/375551}, \href
  {http://adsabs.harvard.edu/abs/2003ApJ...588L.117L} {588, L117}

\bibitem[\protect\citeauthoryear{{Lovis} \& {Fischer}}{{Lovis} \&
  {Fischer}}{2010}]{Lovis10}
{Lovis} C.,  {Fischer} D.,  2010, {Radial Velocity Techniques for Exoplanets}.
pp 27--53

\bibitem[\protect\citeauthoryear{{McQuillan}, {Mazeh}  \&
  {Aigrain}}{{McQuillan} et~al.}{2014}]{McQuillan14}
{McQuillan} A.,  {Mazeh} T.,   {Aigrain} S.,  2014, \mn@doi [The Astrophysical
  Journal Supplement Series] {10.1088/0067-0049/211/2/24}, \href
  {https://ui.adsabs.harvard.edu/\#abs/2014ApJS..211...24M} {211, 24}

\bibitem[\protect\citeauthoryear{{Morris}}{{Morris}}{1985}]{Morris85}
{Morris} S.~L.,  1985, \mn@doi [\apj] {10.1086/163359}, \href
  {http://adsabs.harvard.edu/abs/1985ApJ...295..143M} {295, 143}

\bibitem[\protect\citeauthoryear{{Morris} \& {Naftilan}}{{Morris} \&
  {Naftilan}}{1993}]{Morris93}
{Morris} S.~L.,  {Naftilan} S.~A.,  1993, \mn@doi [\apj] {10.1086/173488},
  \href {https://ui.adsabs.harvard.edu/#abs/1993ApJ...419..344M} {419, 344}

\bibitem[\protect\citeauthoryear{{Penoyre} \& {Stone}}{{Penoyre} \&
  {Stone}}{2019}]{Penoyre18}
{Penoyre} Z.,  {Stone} N.~C.,  2019, \mn@doi [\aj] {10.3847/1538-3881/aaf965},
  \href {https://ui.adsabs.harvard.edu/abs/2019AJ....157...60P} {157, 60}

\bibitem[\protect\citeauthoryear{{Pfahl}, {Arras}  \& {Paxton}}{{Pfahl}
  et~al.}{2008}]{Pfahl08}
{Pfahl} E.,  {Arras} P.,   {Paxton} B.,  2008, \mn@doi [\apj] {10.1086/586878},
  \href {http://adsabs.harvard.edu/abs/2008ApJ...679..783P} {679, 783}

\bibitem[\protect\citeauthoryear{{Placek}, {Knuth}  \& {Angerhausen}}{{Placek}
  et~al.}{2014}]{Placek14}
{Placek} B.,  {Knuth} K.~H.,   {Angerhausen} D.,  2014, \mn@doi [\apj]
  {10.1088/0004-637X/795/2/112}, \href
  {http://adsabs.harvard.edu/abs/2014ApJ...795..112P} {795, 112}

\bibitem[\protect\citeauthoryear{{Press} \& {Rybicki}}{{Press} \&
  {Rybicki}}{1989}]{Press89}
{Press} W.~H.,  {Rybicki} G.~B.,  1989, \mn@doi [\apj] {10.1086/167197}, \href
  {http://adsabs.harvard.edu/abs/1989ApJ...338..277P} {338, 277}

\bibitem[\protect\citeauthoryear{{Ricker} et~al.,}{{Ricker}
  et~al.}{2015}]{Ricker15}
{Ricker} G.~R.,  et~al., 2015, \mn@doi [Journal of Astronomical Telescopes,
  Instruments, and Systems] {10.1117/1.JATIS.1.1.014003}, \href
  {http://adsabs.harvard.edu/abs/2015JATIS...1a4003R} {1, 014003}

\bibitem[\protect\citeauthoryear{{Seager}}{{Seager}}{2010}]{Seager10}
{Seager} S.,  2010, {Exoplanet Atmospheres: Physical Processes}

\bibitem[\protect\citeauthoryear{{Seager} \& {Mall{\'e}n-Ornelas}}{{Seager} \&
  {Mall{\'e}n-Ornelas}}{2003}]{Seager03}
{Seager} S.,  {Mall{\'e}n-Ornelas} G.,  2003, \mn@doi [\apj] {10.1086/346105},
  \href {http://adsabs.harvard.edu/abs/2003ApJ...585.1038S} {585, 1038}

\bibitem[\protect\citeauthoryear{{Shporer}}{{Shporer}}{2017}]{Shporer17}
{Shporer} A.,  2017, \mn@doi [Publications of the Astronomical Society of the
  Pacific] {10.1088/1538-3873/aa7112}, \href
  {https://ui.adsabs.harvard.edu/\#abs/2017PASP..129g2001S} {129, 072001}

\bibitem[\protect\citeauthoryear{{Stassun}, {Collins}  \& {Gaudi}}{{Stassun}
  et~al.}{2017}]{Stassun17}
{Stassun} K.~G.,  {Collins} K.~A.,   {Gaudi} B.~S.,  2017, \mn@doi [\aj]
  {10.3847/1538-3881/aa5df3}, \href
  {http://adsabs.harvard.edu/abs/2017AJ....153..136S} {153, 136}

\bibitem[\protect\citeauthoryear{{Stassun} et~al.,}{{Stassun}
  et~al.}{2018}]{Stassun18}
{Stassun} K.~G.,  et~al., 2018, \mn@doi [\aj] {10.3847/1538-3881/aad050}, \href
  {https://ui.adsabs.harvard.edu/abs/2018AJ....156..102S} {156, 102}

\bibitem[\protect\citeauthoryear{{VanderPlas}}{{VanderPlas}}{2017}]{Vanderplas17}
{VanderPlas} J.~T.,  2017, preprint, \href
  {http://adsabs.harvard.edu/abs/2017arXiv170309824V} {} (\mn@eprint {arXiv}
  {1703.09824})

\bibitem[\protect\citeauthoryear{{Welsh}, {Orosz}, {Seager}, {Fortney},
  {Jenkins}, {Rowe}, {Koch}  \& {Borucki}}{{Welsh} et~al.}{2010}]{Welsh10}
{Welsh} W.~F.,  {Orosz} J.~A.,  {Seager} S.,  {Fortney} J.~J.,  {Jenkins} J.,
  {Rowe} J.~F.,  {Koch} D.,   {Borucki} W.~J.,  2010, \mn@doi [\apjl]
  {10.1088/2041-8205/713/2/L145}, \href
  {http://adsabs.harvard.edu/abs/2010ApJ...713L.145W} {713, L145}

\bibitem[\protect\citeauthoryear{{Winn} \& {Fabrycky}}{{Winn} \&
  {Fabrycky}}{2015}]{Winn15}
{Winn} J.~N.,  {Fabrycky} D.~C.,  2015, \mn@doi [\araa]
  {10.1146/annurev-astro-082214-122246}, \href
  {http://adsabs.harvard.edu/abs/2015ARA%26A..53..409W} {53, 409}

\bibitem[\protect\citeauthoryear{{Wittenmyer} et~al.,}{{Wittenmyer}
  et~al.}{2017}]{Wittenmyer17}
{Wittenmyer} R.~A.,  et~al., 2017, \mn@doi [\aj] {10.3847/1538-3881/aa9894},
  \href {https://ui.adsabs.harvard.edu/\#abs/2017AJ....154..274W} {154, 274}

\bibitem[\protect\citeauthoryear{{de Wit} et~al.,}{{de Wit}
  et~al.}{2017}]{deWit17}
{de Wit} J.,  et~al., 2017, \mn@doi [\apjl] {10.3847/2041-8213/836/2/L17},
  \href {http://adsabs.harvard.edu/abs/2017ApJ...836L..17D} {836, L17}

\makeatother
\end{thebibliography}
\bsp

\appendix

\section{The \texttt{OoT} package}
\label{code}

Alongside this paper, we present a public python code, \texttt{OoT} (short for \textit{Out-of-Transit}), for calculating light curves and radial velocity profiles for planetary systems without transits. It can be found at \url{https://github.com/zpenoyre/OoT}.

The light curves are calculated using equations \ref{deltaTides}, \ref{deltaBeaming} and \ref{deltaReflection}, and they thus include the effects of tides, beaming and reflections. Radial velocity profiles include the effects of orbital motion and tides (see PS18, equations 74 and 77).

Though we focus on out-of-transit effects, all calculations are valid for all orientations, including those in which the planet eclipses the star. If a user wishes to model a transit as well, they can use for example the \texttt{BATMAN} package \citep{Kreidberg15}. \texttt{OoT} contains a function which will convert the parameters of the system to those required by \texttt{BATMAN} (with the exception of the limb darkening parameters).

The user also has the option to model secondary eclipses, under the assumption that the planet acts as a Lambert sphere. A Lambert sphere, when seen from the direction of illumination, is uniform in surface brightness (the full moon is an excellent example of this). Thus we calculate the area of the planet blocked by the star at any given time (which is simply the area of intersection of two circles), and we have an excellent approximation to the effects of the star eclipsing the planet.

The system properties which must be supplied, and their units, are detailed in table \ref{codeParameters}.

As the equations governing out-of-transit behaviours are simple and analytic, the calculation is very efficient. The only bottleneck comes from the numerical solution of $\eta$ as a function of $t$ (from equation \ref{tEta}). We apply an efficient and accurate approximation to make this directly calculable.

Let $\eta_0$ satisfy
\begin{equation}
\eta_0=\sqrt{\frac{GM}{a^3}} t.
\end{equation}
Now assume $\eta = \eta_0 + \eta_1$ where $\eta_1 \ll \eta_0$. We can substitute this back into equation \ref{tEta} and subtract all terms involving $t$ to give
\begin{equation}
\eta_1 = e \sin(\eta_0+\eta_1) = e \sin\eta_0 + O(e^2).
\end{equation}
One can repeat the same exercise to find the second- and third-order components (and indeed we could go to infinity but we show some restraint here) to give
\begin{equation}
\eta = \eta_0 + \eta_1 + \eta_2 +\eta_3 + O(e^4)
\end{equation}
where
\begin{equation}
\eta_2 = e^2 \sin \eta_0 \cos \eta_0
\end{equation}
and
\begin{equation}
\eta_3 = e^3 \sin \eta_0 \left( 1 - \frac{3 \sin^2 \eta_0}{2} \right).
\end{equation}

Thus $\eta(t)$ can be found directly to sufficient accuracy, meaning all formulae necessary for calculating out-of-transit behaviour can be computed effectively instantaneously. Indeed, input times can be given to \texttt{OoT} as an array and the calculation is vectorised, thus the computational cost should not scale. It is also possible to revert back to the exact solution, although this is significantly slower.

\begin{figure}
\begin{centering}
\includegraphics[width=0.45\textwidth]{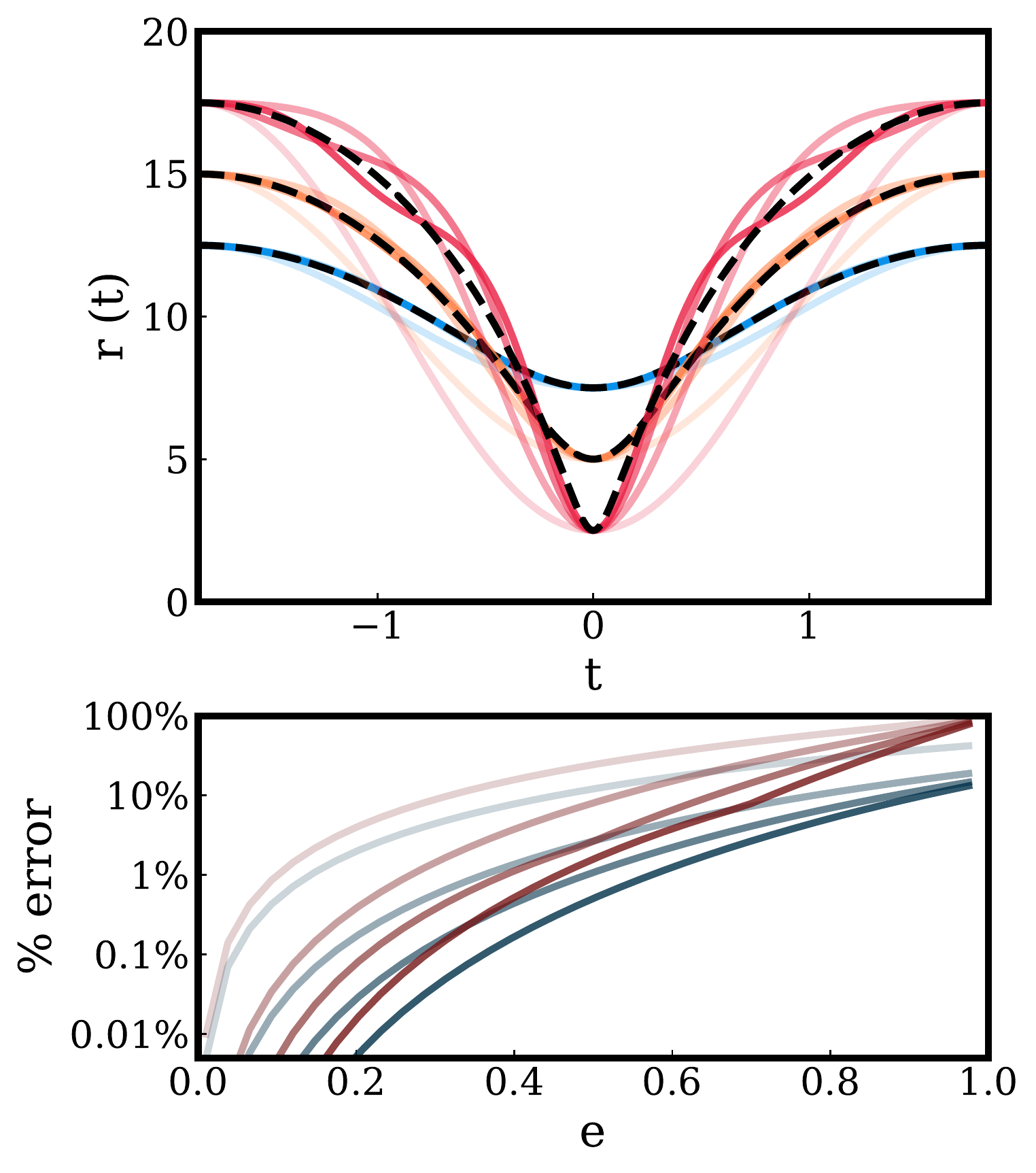}
\caption{Upper panel: Exact orbital radius (dashed black line) compared to the approximate solutions accurate to increasing powers of $e$ (darker lines). The orbits shown have $e=0.25$ (blue), $e=0.5$ (orange) and $e=0.75$ (red). Lower panel: The maximum (dark red) and time averaged (dark blue) fractional error in the radius derived from our approximate solution over one period. Darker lines show higher orders of the approximation.}

\label{approximation}
\end{centering}
\end{figure}

Figure \ref{approximation} shows the accuracy of this approximation over one orbital period for a range of eccentricities. It can be seen that it is generally a good fit, though less so for very high $e$, and that the largest error comes from the approximate solution dawdling too long near periapse. Note that although our most accurate solution only contains terms $\propto e^3$, the radius depends on $\eta$ only through a term $e \cos \eta$ and thus the radius is accurate up to terms of order approximately $e^6$.

The code is written entirely in units of days, $M_\odot$, and $R_\odot$, save for velocities, which are always returned in $m s^{-1}$. We have included functionality to convert units to years, seconds, Jupiter- and Earth-radii and masses (for $R_p$ and $M_p$), and astronomical units (for $a$).

\begin{table}
\centering
\begin{tabular}{ c| c} 
Parameter & Default Unit \\
 \hline
 $M$ - stellar mass & $M_\odot$  \\
 $M_p$ - planet mass & $M_\odot$ ($\approx 1000 M_j$) \\
 $R$ - stellar radius & $R_\odot$ \\
 $R_p$ - planet radius & $R_\odot$ ($\approx 10 R_j$) \\
 $a$ - semi-major axis & $R_\odot$ ($\approx 0.005 AU$) \\
 $e$ - eccentricity & unitless \\
 $A_g$ - geometric albedo & unitless \\
 $\beta$ - stellar response to tides & unitless \\
 $\theta_v$ - polar viewing angle & radians \\
 $\phi_v$ - azimuthal viewing angle & radians \\
 $t_p$ - time at periapse & days\\
\end{tabular}
\caption{The parameters used by the \texttt{OoT} package to calculate light curves, radial velocity profiles and photometric power spectra. All orbital parameters refer to the planet. Though the user can vary $A_g$ and $\beta$, throughout this paper we have used assumed values (0.15 and 1 respectively).}
\label{codeParameters}
\end{table}

\section{Analytic calculation of power spectra}
\label{analytic}

The power spectra of out-of-transit signals are strongly peaked at harmonics of the orbital frequency. This means that we can calculate the amplitude of each peak by expressing the curves as Fourier series. We may then perform relatively simple (though not necessarily short) calculations to analytically derive the power spectrum, and in doing so explain the form of the power spectra for tides, reflections, and beaming.


\subsection{Preliminaries}

We can decompose some general function $g(x)$, which is periodic (satisfying $g(x)=g(x+2\pi)$), into a Fourier series of the form
\begin{equation}
g(x) = \frac{\alpha_0}{2} + \sum_{n=1}^{\infty} \left[\alpha_n \cos(n x) + \beta_n \sin(n x)\right].
\label{eq:fourier}
\end{equation} 
The constant coefficients obey
\begin{equation}
\alpha_n = \frac{1}{\pi} \int_{-\pi}^{\pi} g(x) \cos(n x) dx
\end{equation}
and
\begin{equation}
\beta_n = \frac{1}{\pi} \int_{-\pi}^{\pi} g(x) \sin(n x) dx.
\end{equation}

In this derivation, we will make use of the orthonormality condition
\begin{equation}
\begin{aligned}
\frac{1}{\pi} \int_{-\pi}^{\pi} \cos(n x) \cos(m x) dx = \begin{cases}
    2,& n=m=0\\
    1,& n=m\ne 0\\
    0,& n\ne m
\end{cases}
\end{aligned}
\end{equation}
(and similarly for sine terms) and of the formulae
\begin{equation}
\cos n x \cos m x  = \frac{\cos[(n+m) x] + \cos[(n-m) x]}{2},
\end{equation}
\begin{equation}
\sin n x \sin m x  = \frac{\cos[(n-m) x] - \cos[(n+m) x]}{2}
\end{equation}
and
\begin{equation}
\sin n x \cos m x  = \frac{\sin[(n+m) x] + \sin[(n-m) x]}{2}.
\end{equation}

The last tool we shall need is the ability to express a product of many sine and cosine terms, of the form $\sin^a x \cos^b x$, as a sum of harmonics, i.e. terms of the form $\sin{nx}$ and $\cos{nx}$. (This will be useful because the harmonics can be easily plugged into the orthonormality condition, above.)

Let $C(n,a,b)$ be coefficient of the $n^{th}$ harmonic in the expansion of $\sin^a x \cos^b x$. By expressing sine and cosine terms in their complex form, it can be shown that
\begin{equation}
\begin{aligned}
C(n,a,b) =& 2^{-(a+b)} (-1)^{int(\frac{a}{2})} \Bigg[
\sum_{j=A_{min}}^{A_{max}} (-1)^j \frac{a! b!}{j! (a-j)! (\frac{a+b-n}{2} - j)! (\frac{n+b-a}{2}+j)!} \\
&+\sum_{j=B_{min}}^{B_{max}} (-1)^{j+a} \frac{a!b!}{j!(a-j)!(\frac{n+a+b}{2}-j)!(\frac{b-a-n}{2}+j)!} \Bigg]
\end{aligned}
\end{equation}
where
\begin{equation}
A_{min} = max\left(0,\frac{a-b-n}{2}\right) \ \mathrm{and} \ A_{max} = min\left(a,\frac{a+b-n}{2}\right),
\end{equation}
\begin{equation}
B_{min} = max\left(0,\frac{a-b+n}{2} \right) \ \mathrm{and} \ B_{max} = min\left(a,\frac{a+b+n}{2}\right)
\end{equation}
and $int(\frac{a}{2})$ is the largest integer less than or equal to $\frac{a}{2}$.

We haven't yet specified whether $C(n,a,b)$ is the coefficient of the sine or cosine term in the expansion, i.e. the coefficient of $\sin{nx}$ or $\cos{nx}$. Firstly it can be seen that no real $C$ exists when $(a+b+n)$ is odd. The value of $a$ determines which terms the expansion can be expressed in:
\begin{equation}
\begin{aligned}
\sin^a x \cos^b x =& \sum_{n=0}^{a+b} C(n,a,b) \cos nx, \ \ \mathrm{\ even} \ a \\
 & \sum_{n=0}^{a+b} C(n,a,b) \sin nx, \ \ \mathrm{\ odd} \ a
\end{aligned}
\end{equation}

When $a$ is even, there are no sine terms in the expansion, and vice versa.

\subsection{Application to light curves}

With reference to Equation~\ref{eq:fourier}, we would like to express the time evolution of a single light curve signal (which we'll denote by a general $x$) as
\begin{equation}
\delta_x(t) = A_0 + \sum_{n=1}^{\infty}\left[ A_n \cos(n \omega t) + B_n \sin(n \omega t)\right].
\label{deltaExpansion}
\end{equation}

We will ignore the constant term $A_0$. To compute the other coefficients $A_n$ and $B_n$ of equation \ref{deltaExpansion}, we must evaluate
\begin{equation}
A_n = \frac{\omega}{\pi} \int_{-\frac{\pi}{\omega}}^{\frac{\pi}{\omega}} \delta_x (t) \cos(n \omega t) dt
\end{equation}
and similar for $B_n$ with sine terms, recalling that $\delta_x(t)$ is periodic at $P =\frac{2\pi}{\omega}$. 

However, it is not trivial to express $\delta_x(t)$ as a simple function of time. Instead it will be more convenient to re-express this integral and all terms within in terms of $\eta$ (defined by Equations \ref{eta} and \ref{tEta}).

Using
\begin{equation}
dt = \frac{1- e \cos \eta}{\omega} d\eta,
\end{equation}
we can write
\begin{equation}
\label{mainIntegral}
A_n = \frac{1}{\pi} \int_{-\pi}^{\pi} (1- e \cos \eta)  \ \cos (n \omega t) \ \delta_x(\eta) \ d\eta,
\end{equation}
and similar for $B_n$ with sine terms.

In order to evaluate equation \ref{mainIntegral} (and its $B_n$ counterpart), we need to express each individual piece in terms of functions of $\sin{n\eta}$ and $\cos{n\eta}$, for integer $n$. We can then use the orthonormality condition to cancel most terms.

\subsubsection{Expansion of $\cos(n\omega t)$}

Let us start by expanding $\cos(n\omega t)$ and $\sin(n\omega t)$ by rewriting Equation \ref{tEta} as
\begin{equation}
\omega t = \eta - e \sin \eta.
\end{equation}

Using the fact that $e \sin \eta < \eta$ at all times (as $e<1$) we can perform a Taylor expansion around $\omega t = \eta$. Doing so yields

\begin{equation}
\cos(n\omega t) =  \sum_{j=0}^{\infty} (-1)^j \left( \frac{(n e \sin \eta)^{2j}}{(2j)!} \cos n\eta + \frac{(n e \sin \eta)^{2j+1}}{(2j+1)!} \sin n\eta \right)
\label{eq:cosSeries}
\end{equation}
and similarly
\begin{equation}
\sin(n\omega t) =  \sum_{j=0}^{\infty} (-1)^j \left( \frac{(n e \sin \eta)^{2j}}{(2j)!} \sin n\eta - \frac{(n e \sin \eta)^{2j+1}}{(2j+1)!} \cos n\eta \right).
\label{eq:sinSeries}
\end{equation}

These equations can be re-expressed using $C(n,a,b)$ as
\begin{equation}
\begin{aligned}
\cos(n\omega t) =&  \sum_{j=0}^{\infty} \sum_{k=0}^{j} (-1)^j \Bigg( C(2k,2j,0) \frac{(n e)^{2j}}{(2j)!} \frac{\cos (2k+n)\eta + \cos (2k-n)\eta}{2} \\+& C(2k+1,2j,0) \frac{(n e)^{2j+1}}{(2j+1)!}  \frac{\cos (2k+1+n)\eta - \cos (2k+1-n)\eta}{2} \Bigg)
\end{aligned}
\end{equation}
and
\begin{equation}
\begin{aligned}
\sin(n\omega t) =&  \sum_{j=0}^{\infty} \sum_{k=0}^{j} (-1)^j \Bigg( C(2k,2j,0) \frac{(n e)^{2j}}{(2j)!} \frac{\sin (2k+n)\eta + \sin (2k-n)\eta}{2} \\-& C(2k+1,2j,0) \frac{(n e)^{2j+1}}{(2j+1)!} \frac{\sin (2k+1+n)\eta + \sin (2k+1-n)\eta}{2} \Bigg).
\end{aligned}
\end{equation}

\subsubsection{Expansion of $\delta$}

Whether for tides, beaming or reflections, $\delta_x$ contains trigonometric functions of $\Phi$ (often re-expressed in terms of $\psi_v = \phi_v - \Phi$).

We can rearrange Equation \ref{eta} to give
\begin{equation}
\cos \Phi = \frac{\cos \eta - e}{1 - e \cos \eta}
\label{eq:cosPhi}
\end{equation}

and

\begin{equation}
\sin \Phi = \frac{\sqrt{1-e^2} \sin \eta}{1 - e \cos \eta}.
\label{eq:sinPhi}
\end{equation}

For tides and reflections, $\delta$ also depends on $r(t) = a(1-e\cos{\eta})$, which introduces further factors of $\cos{\eta}$.

\subsection{Example: First order correction to tides}
\label{tidesCorrection}

We have presented all the the components needed to calculate a general power in each harmonic for a given eccentricity. What we certainly have not done is worked this through to a formula or numerical value. Instead, we shall attempt to salvage some simple intuition from this tangled nest of summations.

We shall find explicitly the first order correction to the power spectrum due to eccentricity in a system with \textit{only} tidal signatures visible (i.e. where beaming and reflection have comparably negligible amplitudes). To simplify, we will further choose $\phi_v = 0$ and $\theta_v = \frac{\pi}{2}$.

To first order in $e$, Equations~\ref{eq:cosSeries} and~\ref{eq:sinSeries} for the expansion of $\cos{(n\omega t)}$ and $\sin{(n\omega t)}$ reduce to 
\begin{equation}
\cos(n\omega t) = \cos n \eta - e \frac{n}{2}\big( \cos\left[(n+1)\eta\right] - \cos\left[(n-1)\eta\right] \big) + O(e^2)
\end{equation}

and
\begin{equation}
\sin(n\omega t) = \sin n \eta - e \frac{n}{2}\big( \sin\left[(n+1)\eta\right] + \sin\left[(n-1)\eta\right] \big) + O(e^2).
\end{equation}

Meanwhile, to first order in $e$, Equations~\ref{eq:cosPhi} and~\ref{eq:sinPhi} for the expansion of $\Phi$ reduce to
\begin{equation}
\cos \Phi =\cos \eta - e \sin^2 \eta + O(e^2)
\end{equation}

and
\begin{equation}
\sin \Phi =\sin \eta + e \sin \eta \cos \eta+ O(e^2).
\end{equation}



Let us express $\delta_t$ as:
\begin{equation}
\delta_t = \kappa \frac{3 \cos^2 \Phi - 1 }{(1- e\cos \eta)^3}, \\
\end{equation}

To first order in $e$, then,

\begin{equation}
\delta_t = \frac{\kappa}{2} \left( 1 + 3 \cos 2\eta + \frac{3 e}{2} (5 \cos 3\eta + 3 \cos \eta)\right) + O(e^2)
\end{equation}
where the constant $\kappa$ encodes all other parameters of the system.

\subsubsection{Coefficients $A_n$}

By choosing $\phi_v = 0$ we ensure the light curve is an even function, so only the cosine terms in equation \ref{deltaExpansion} remain.

Putting the pieces in Equation~\ref{mainIntegral} together, we find that in the circular ($e=0$) case, $A_2 =\frac{3\kappa}{2}$ and all other terms are 0.

For the non-circular case, $A_1 = \frac{3}{4}e \kappa$ and $A_3 = \frac{21}{4} e \kappa$.

Thus the ratio of amplitude in the third (first) harmonic to that in the second is $\frac{7 e}{2}$ ($\frac{e}{2}$). As expected, the height of the harmonics scales as $e$.

There is scope for deriving the full range of more complex behaviours for larger $e$, arbitrary viewing angles and other periodic fluctuations. However, given the length of this derivation we believe this is left to the (very) interested reader in their (very) spare time.

\label{lastpage}

\end{document}